\documentclass[11pt]{article}
\pdfoutput=1
\usepackage[centertags]{amsmath}
\usepackage[square, comma, sort&compress,numbers]{natbib}
\usepackage{array,multirow}
\numberwithin{equation}{section}
\usepackage{amssymb}
\usepackage{graphicx}
\usepackage{color}
\usepackage{setspace}
 
\usepackage{epsfig}
\def\half{{1 \over 2}}

\def\Or[#1]{{\text{O}}\left({#1}\right)}
\def\dotl[#1,#2]{\left\langle #1, #2 \right\rangle}
\def\dotlb[#1,#2]{[ #1, #2 ]}
\def\dotp[#1,#2]{(#1) \cdot (#2)}
\def\aff[#1,#2]{\hat{#1}(#2)}
\def\n4sym{{\cal N}=4 SYM}
\def\>{\rangle}
\def\<{\langle}
\def\weight[#1,#2,#3]{\{(#1),#2,#3\}}
\def\ads[#1]{$\text{AdS}_{#1}$}

\newcommand{\ba}{\begin{eqnarray}}
\newcommand{\ea}{\end{eqnarray}}
\newcommand{\be}{\begin{eqnarray}}
\newcommand{\ee}{\end{eqnarray}}
\newcommand{\bq}{\begin{equation}}
\newcommand{\eq}{\end{equation}}
\newcommand{\benn}{\begin{equation*}}
\newcommand{\eenn}{\end{equation*}}
\newcommand{\bi}{\begin{itemize}}  
\newcommand{\ei}{\end{itemize}}

\newcommand{\CO}{{\cal O}}

\newcommand{\CV}{{\cal V}}
\newcommand{\nn}{\nonumber}

\newcommand\oo\infty
\newcommand\s\sigma
\newcommand\de\delta
\newcommand\De\Delta

\newcommand\f\phi
\newcommand\g\gamma
\newcommand\x\times

\newcommand{\ra}{\rightarrow}
\newcommand{\lra}{\leftrightarrow}
\newcommand{\fr}{\frac}
\newcommand{\comm}[2]{[#1,#2]}

\newcommand{\Ocal}{{\cal O}}
\newcommand{\Dcal}{{\cal D}}
\newcommand{\Ncal}{{\cal N}}
\newcommand{\Mcal}{{\cal M}}
\newcommand{\Pcal}{{\cal P}}
\newcommand{\Vcal}{{\cal V}}

\newcommand\G{\Gamma}

\usepackage{jheppub}
\usepackage{comment}

\makeatletter
\def\@fpheader{\vspace{-.1cm}}
\makeatother

\title{Eikonalization of Conformal Blocks}
\author[a,b]{A.\ Liam Fitzpatrick,}
\author[c]{Jared Kaplan,}
\author[d]{Matthew T.\ Walters,}
\author[c]{and Junpu Wang}
\affiliation[a]{Stanford Institute for Theoretical Physics, Stanford University, \\
Via Pueblo, Stanford, CA 94305, U.S.A.}
\affiliation[b]{SLAC National Accelerator Laboratory, \\
Sand Hill Road, Menlo Park, CA 94025, U.S.A.}
\affiliation[c]{Department of Physics and Astronomy, Johns Hopkins University, \\
Charles Street, Baltimore, MD 21218, U.S.A.}
\affiliation[d]{Department of Physics, Boston University, \\
Commonwealth Avenue, Boston, MA 02215, U.S.A.}
\emailAdd{fitzpatr@stanford.edu}
\emailAdd{jaredk@pha.jhu.edu}
\emailAdd{mtwalter@bu.edu}
\emailAdd{jwang217@jhu.edu}
\abstract{
Classical field configurations such as the Coulomb potential and Schwarzschild solution are built from the t-channel exchange of many light degrees of freedom.  We study the CFT analog of this phenomenon, which we term the `eikonalization' of conformal blocks.  We show that when an operator $T$ appears in the OPE $\CO(x)\CO(0)$, then the large spin Fock space states $[TT \cdots T]_{\ell}$ also appear in this OPE with a computable coefficient.  The sum over the exchange of these Fock space states in an $\< \CO \CO \CO \CO \>$ correlator build the classical `$T$ field' in the dual AdS description.  In some limits the sum of all Fock space exchanges can be represented as the exponential of a single $T$ exchange in the 4-pt correlator of $\CO$.   Our results should be useful for systematizing $1/\ell$ perturbation theory in general CFTs and simplifying the computation of large spin OPE coefficients.  As examples we obtain the leading $\log \ell$ dependence of Fock space conformal block coefficients, and we directly compute the OPE coefficients of the simplest `triple-trace' operators. 
}
\arxivnumber{}

\begin{document}

\maketitle
\flushbottom

%
%

\section{Introduction and Summary}

The correlation functions of local operators in Conformal Field Theories (CFTs) must satisfy fundamental consistency conditions encoding conformal symmetry and quantum mechanical unitarity.  In the bootstrap approach, one attempts to constrain or compute the CFT correlators, or equivalently, the CFT spectrum and operator product expansion (OPE) coefficients, using only these fundamental principles as an input.  The bootstrap, which was very successful in two dimensions \cite{FerraraOriginalBootstrap1,PolyakovOriginalBootstrap2}, has recently yielded powerful numerical  \cite{Rattazzi:2008pe,Rychkov:2009ij,Caracciolo:2009bx,Poland:2010wg,Rattazzi:2010gj,Rattazzi:2010yc,Vichi:2011ux,Poland:2011ey,Rychkov:2011et,ElShowk:2012ht,Liendo:2012hy,ElShowk:2012hu, Beem:2013qxa,Kos:2013tga,Gliozzi:2013ysa,El-Showk:2013nia,Gaiotto:2013nva,Bashkirov:2013vya, El-Showk:2014dwa, Kos:2014bka, Beem:2014zpa, Chester:2014gqa, Chester:2014fya, Simmons-Duffin:2015qma, Paulos:2014vya, Bobev:2015vsa, Bobev:2015jxa} and analytical \cite{JP, Hellerman:2009bu, ElShowk:2011ag, Fitzpatrick:2012yx, KomargodskiZhiboedov, Fitzpatrick:2014vua, Alday:2013cwa, Hartman:2014oaa, Alday:2014tsa, Alday:2015eya, Jackson:2014nla, Kaviraj:2015cxa, Kaviraj:2015xsa} results.  It is natural to ask how far we can go using analytical techniques and only a smattering of CFT data.

To address this question we need not grasp about in the dark, because the AdS/CFT correspondence suggests specific expectations.  Long-distance locality in AdS, and the existence of universal long-range forces (such as gravity) both lead to predictions for the spectrum and OPE coefficients of the CFT.  In recent work \cite{Fitzpatrick:2012yx, KomargodskiZhiboedov, Fitzpatrick:2014vua} these predictions have been derived from the bootstrap, without any reference to AdS, and for all unitary CFTs in $d \geq 3$ dimensions, with more intricate and powerful results in $d=2$ at large central charge. 

CFTs have a Fock space of states at large spin $\ell$, corresponding to a physical Fock space of well-separated objects in AdS.  Since the AdS Hamiltonian is the dilatation operator of the CFT, the anomalous dimensions $\gamma(\ell)$ of these states represent AdS interaction energies between distant objects.  The $\gamma(\ell)$ are determined by OPE coefficients with low-twist operators, corresponding to couplings between AdS objects and light (or low mass) fields \cite{Fitzpatrick:2014vua}.   For example, the exchange of the stress energy tensor $T_{\mu \nu}$ in the CFT roughly corresponds with the exchange of a virtual graviton in AdS, and the universality of $T_{\mu \nu}$ OPE coefficients leads to the equivalence principle in AdS.  

\begin{figure}[t!]
\begin{center}
\includegraphics[width=0.85\textwidth]{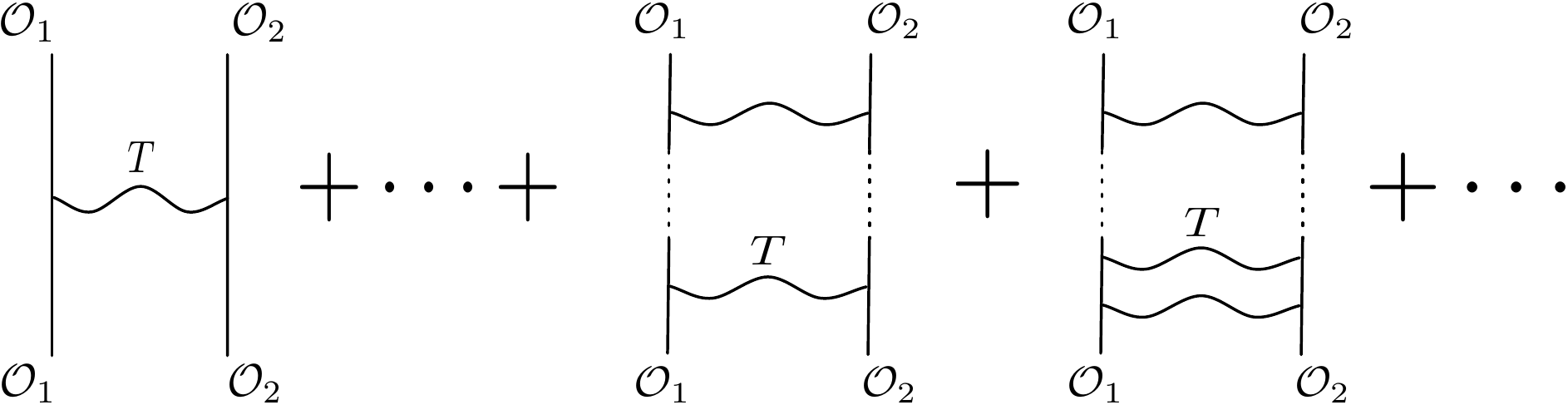}
\caption{ This figure indicates how one might sum over multiple virtual exchanges in order to construct an effective classical background.  We would like to understand this process directly in the CFT, with minimal assumptions.  When the first diagram determines the sum of the rest, we say that the conformal blocks `eikonalize'.  }
 \label{fig:SumOverTExchange} 
\end{center}
\end{figure}

We would like to study corrections to these results, specifically the summation\footnote{Although we use the word eikonalization, it should be noted that we are not studying the traditional eikonal limit of high energy and large impact parameter, but of fixed energy and large impact parameter, where impact parameter grows with spin.} or `eikonalization' of multiple virtual exchanges into an effective classical background \cite{Fitzpatrick:2015zha}, as pictured in figure \ref{fig:SumOverTExchange}.  We will see that these effects are also essentially universal, but first we will run into an obstruction.  In the process of overcoming it we will make connections with Mellin amplitude \cite{Mack, Macksummary, JoaoMellin, NaturalLanguage, Paulos:2011ie} asymptotics \cite{AdSfromCFT} and an elementary theorem of Darboux \cite{Dingle, Braaksma:1997:DTS:256442.256449}, which justify a simple and general procedure for extracting the OPE coefficient of any large spin operator.

\subsection*{A CFT Sandbox}

To discuss the details, we will be making extensive use of the idea of conformal blocks \cite{Dolan:2000ut, Dolan:2003hv, Dolan:2011dv}, also known as conformal partial waves, and the CFT bootstrap equation.  These were briefly reviewed in a relevant context in \cite{Fitzpatrick:2014vua} and in many other recent works.  We seek to understand if the conformal partial waves associated with the exchange of a full Fock space can be resummed or `eikonalized' into a simple closed form.  Directly on the AdS side, these issues have been explored \cite{Polchinski:2002jw, Joaothesis, JoaoBFKL, JoaoRegge} at high energy with fixed impact parameter, leading to an AdS version of the eikonal limit.  In the case of CFT$_2$ the resummation of stress tensor \cite{Fitzpatrick:2014vua} and current exchange \cite{Fitzpatrick:2015zha} have already been observed, but we will see that the general story is more subtle.

\begin{figure}[t!]
\begin{center}
\includegraphics[width=0.75\textwidth]{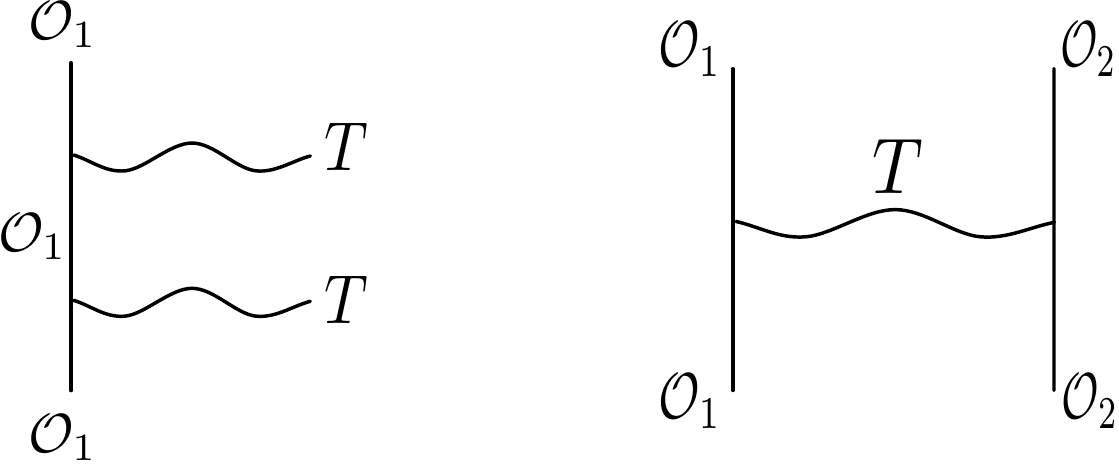}
\caption{ This figure indicates conformal partial waves that necessarily contribute to two different 4-pt CFT correlators, based on the assumed OPEs.  We indicate the conformal block on the left as $\CO_1 T \to \CO_1 \to \CO_1 T$.
}
 \label{fig:TwoNecessaryBlocks} 
\end{center}
\end{figure}

For our purposes it will be sufficient to study just a few primary operators in a general CFT$_d$, which we refer to as 
\be \label{eq:AssumedCFTData}
\CO_1, \ \CO_2, \ T  \ \ \ \mathrm{with}  \ \ \ \CO_i(x) \CO_i(0)  \supset T
\ee  
where by $\supset$ we mean `is included in the OPE'.  We use $\Delta_1, \Delta_2, \Delta_T$ to refer to the dimensions of these operators, and $\tau_i$ and $\tau_T$ to refer to their twists $\tau \equiv \Delta - \ell$.  We will be thinking of $T$ as a low dimension or `light' operator, such as the stress tensor, and $\CO_i$ as heavier sources.  The indicated OPE immediately implies that certain specific conformal partial waves must contribute to correlators such as $\< \CO_1 \CO_1 T T \>$ and $\< \CO_1 \CO_1 \CO_2 \CO_2 \>$, as pictured in figure \ref{fig:TwoNecessaryBlocks}.  

The theorem \cite{AldayMaldacena, Fitzpatrick:2012yx, KomargodskiZhiboedov, Fitzpatrick:2014vua} referred to above states that \emph{in the OPE $A(x) B(0)$ of any two primary operators there exist new primaries $[AB]_{n, \ell}$ labeled by positive integers $n, \ell$ at large $\ell$, with dimension $\tau_A + \tau_B + 2n + \ell + \gamma(n, \ell)$, where the anomalous dimension $\gamma(n, \ell) \to 0$ as $\ell \to \infty$ at a prescribed power-law rate.}  This immediately implies the existence of operators
\be
[\CO_1 \CO_2]_{n, \ell},  \ [\CO_1 T]_{n, \ell},  \ [T T]_{n, \ell}, \ \cdots, \ \left[ [\CO_1 \CO_1]_{n, \ell} T \right]_{n', \ell'}, \cdots
\ee
for all possible combinations at large $\ell$.  We will sometimes abuse AdS/CFT language and refer to $[AB]_{n, \ell}$ as `double-trace' operators, although all of our results are wholly independent of large $N$.  

\begin{figure}[t!]
\begin{center}
\includegraphics[width=0.85\textwidth]{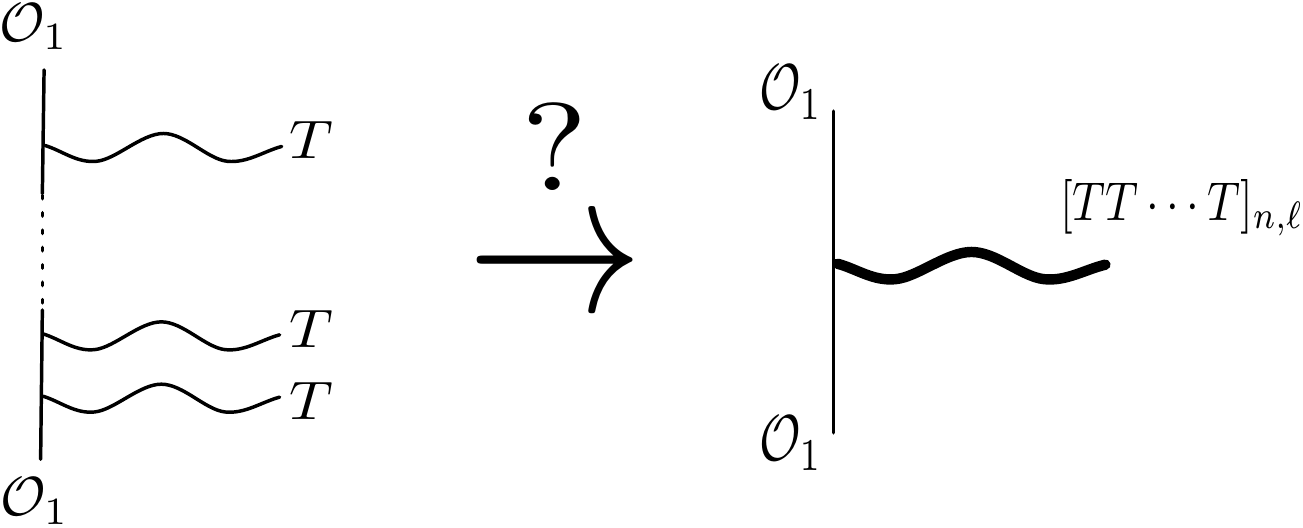}
\caption{ Large $\ell$ Fock space operators should have a universal behavior. This figure indicates how one might try to use known OPE coefficients to construct conformal blocks, and then take the OPE limit in a different channel to obtain new information about  general Fock space states.   By making a simple assumption about the CFT correlators in Mellin space, these OPE limits can to be shown to exist, and give a universal result for OPE coefficients with Fock space states. }
 \label{fig:FromBlocksToOPE} 
\end{center}
\end{figure}

Applying the theorem recursively leads to a Fock space with any number of $\CO_1, \CO_2, $ and $T$, which can be interpreted as a collection of any number of well-separated objects in AdS.  We would like to understand what predictions can be made for the OPE coefficients of these operators, because in particular, the OPE coefficients of
\be
\CO_1(x) \CO_1(0) \supset T, \ [TT]_{n, \ell}, \ [TTT]_{n, \ell}, \cdots
\ee
determine how multiple $T$ exchange generates an effective classical background in AdS.  Physically, it seems reasonable to expect that large $\ell$ operators such as $[TT]_{n, \ell}$ have universal OPE coefficients determined by those of $T$, since we can interpret this operator as a pair of $T$ states that have been well-separated in AdS.   We will see that via the process pictured in figure \ref{fig:FromBlocksToOPE} these OPE coefficients are essentially universal, although there is a barrier to be overcome.

Throughout this paper, we will refer to the conformal block decomposition of a four-point function of the general form $\< \Ocal_1(x_1) \Ocal_2(x_2) \Ocal_3(x_3) \Ocal_4(x_4)\>$ in the channel arising from the $\Ocal_1 \times \Ocal_2$ OPE (or equivalently, the $\Ocal_3 \times \Ocal_4$ OPE)  as the 
\be
\Ocal_1 \Ocal_2 \rightarrow \Ocal_3 \Ocal_4
\ee
 channel, and the conformal block for an exchange of an operator $\CO$ in this channel as the
 \be
 \Ocal_1 \Ocal_2 \rightarrow \CO \rightarrow \Ocal_3 \Ocal_4
 \ee
  conformal block. 

\subsection*{Mellin Amplitude Asymptotics, Darboux's Theorem, and the OPE}

The single conformal block pictured on the left in figure \ref{fig:TwoNecessaryBlocks} cannot be expanded as a sum of conformal blocks in the cross-channel while satisfying unitarity constraints.  Relatedly, the OPE limit of $T(z)T(0)$ in this conformal block will not be well-behaved.  This presents a problem because one might have used an expansion in this limit to compute the OPE coefficients
\be
\< \CO_1 \CO_1 [TT]_{n, \ell} \>
\ee
which must exist at large $\ell$. 
But this procedure will not be well-defined!

The problem is that the conformal block for the correlator
\be
G_{\CO_1}(z) = \< \CO_1(\infty)  T(0) \left( \sum_{\CO_1 \ \mathrm{desc}} | \alpha \> \< \alpha | \right) \CO_1(1) T(z) \>
\ee
has branch cuts in the limit that $z \to 0$ that cannot be interpreted (in a unitary theory) as a sum of operators in the $T(z)T(0)$ OPE.  However, let us write this conformal block in terms of the toy Mellin integral
\be
G_{\CO_1}(z) = \int_{-i \infty}^{i \infty} d \delta \, \Mcal(\delta) \, (1-z)^{-\delta}.
\ee
If we assume that $\Mcal(\delta)$ vanishes exponentially as $\delta \to \pm i \infty$ then $G_{\CO_1}(z)$ will be analytic around $z=0$.  The individual conformal blocks violate this assumption, and so they do not have a good OPE limit in the cross-channel (where $z \to 0$).  However, if the full correlator satisfies this exponential bound, then the OPE limit will exist.  We can still account for the individual contributions from specific conformal blocks by looking at their poles, whose position and residue cannot be altered due to the constraints of conformal symmetry and unitarity.  This reasoning ties the Mellin amplitude asymptotics to the existence of the OPE and the universality of Fock space OPE coefficients.

The large spin OPE coefficients can be obtained\footnote{We should emphasize that specific OPE coefficients for individual $\ell$ cannot be rigorously determined; only their large $\ell$ sums can be computed.  But for simplicity we will talk about `OPE coefficients' as if they take values equal to their averages at large $\ell$.  More rigorous bounds on integrated OPE coefficients can be obtained using the Hardy-Littlewood Tauberian theorem  \cite{Pappadopulo:2012jk, Fitzpatrick:2012yx, Fitzpatrick:2014vua}.} by applying an elementary observation of Darboux \cite{Dingle, Braaksma:1997:DTS:256442.256449}, that the large order behavior of the series expansion of a function $f(z)$ is dominated by the singularities of $f(z)$ on the circle bounding the region of convergence.  In the context of the correlator $\< \CO_1(\infty) \CO_1(1) T(z) T(0) \>$, this means that the large order expansion of the $T(z) T(0)$ OPE must be governed by the $T(z) \CO_1(1)$ OPE.  Since the large spin operators $[TT]_{n, \ell}$ appear at high orders in the $T(z) T(0)$ OPE, their OPE coefficients must be governed by singularities in the $z \to 1$ limit.  These are encoded in the leading poles and residues of the Mellin amplitude, which are independent of the contributions that fix its asymptotic behavior.  

\subsection*{Summary of Results}

In this work we will mostly `follow our nose', using only the assumptions in equation (\ref{eq:AssumedCFTData}), the OPE, the CFT bootstrap, and later on, some reasoning motivated by Darboux's theorem.  But the analysis itself will become a bit technical, so for the casual reader we summarize our results here:
\begin{itemize}
\item A universal part of the large $\ell$ OPE coefficients of operators like $[TT]_{n, \ell}, [TTT]_{n_i, \ell}, \cdots$ are determined by the OPE coefficient of $T$ with $\CO_i(x) \CO_i(0)$.  The exchange of these operators between $\CO_1 \CO_1 \to \CO_2 \CO_2$ can be rewritten in the exponentiated form $e^{P_T g_T(u, v)}$ when $\Delta_i \gg \tau_T$.

\item  The non-universal behavior of the $[TT]_{n,\ell}$ OPE coefficients dictates the existence of other operators in the $\CO_1(x) \CO_2(0)$ OPE.  Conversely, we can compute corrections to the $[TT]_{n,\ell}$ OPE coefficients based on the $\CO_i(x) \CO'(0) \supset T$ OPE for $\tau_{\CO'} \leq \tau_{\CO_i}$.

\item The OPE coefficients of general large spin Fock space operators $[TS]_{n, \ell}$ with $\CO(x) \CO(0)$ can be obtained by applying differential operators or the conglomeration procedure \cite{JoaoMellin, Unitarity} to the singular parts of $\< \CO(\infty) \CO(1) T(z) S(0) \>$ as $z \to 1$.  As an example, we compute the leading OPE coefficients of minimal twist $[TS]_{0, \ell}$ from the $\CO T \to \CO' \to \CO S$ conformal block in section \ref{sec:UniversalityandDarboux}.
\item If one further assumes that the $\< \CO_i \CO_i T \>$ OPE coefficient is proportional to a perturbative parameter `$1/N$', then one can show that it is the contribution of `double-trace' $[\CO T]_{n, \ell}$ operators \cite{AdSfromCFT} that render the $T(z) T(0)$ OPE well-defined.  
\item At a technical level, we discuss and compare three distinct methods for extracting large spin OPE coefficients in sections \ref{sec:Eikonalization}, \ref{sec:LargeSpinOPEfromBootstrap}, and \ref{sec:MellinBoundedness}, and we compute the $[TTT]_{\ell}$ OPE coefficients directly using differential operators in section \ref{sec:UniversalityandDarboux}.
\end{itemize}

These results specifically hold in the limit $\ell \ra \infty$ with the external dimensions $\De_i$ fixed, which has the AdS interpretation of the exchange of well-separated light mediators between two objects with fixed energy. It is important to note that this setup is distinct from the standard `eikonal limit' of \emph{both} large impact parameter and high energy. In the eikonal limit, the large energy of the two objects allows the exchanged light field to be described by a classical shock wave configuration, which can then be used to calculate AdS scattering amplitudes and derive predictions for the associated CFT correlators \cite{Polchinski:2002jw, Joaothesis, JoaoBFKL, JoaoRegge,CostaEikonal,Cornalba:2006xk,Cornalba:2006xm,Cornalba:2007fs}. The universal contributions to exchanged conformal blocks that we derive here arise instead from the approximate Fock space structure of CFT operators at large spin. We refer to this behavior as the `eikonalization' of conformal blocks because of its exponentiated structure at large $\De_i$, suggesting a similar interpretation in terms of classical background fields in AdS. It would be interesting to explore this interpretation further and connect our results with the standard eikonal limit more directly in future work.

We will conventionally write 4-pt correlators using the parametrization
\be
\<  \CO_i(\infty) \CO_i(1) T(z, \bar z) T(0)  \>, \ \ \ \< \CO_1(\infty) \CO_1(1) \CO_2(z, \bar z) \CO_2(0)  \>
\ee
which we note for uniformity of presentation, so that it is clear which OPE limits are obtained by $z \to 0$ versus $z \to 1$.  We refer to $z, \bar z \to 0$ or $z, \bar z \to 1$ as OPE limits, while $\bar z \to 0$ or $1$ with fixed $z$ is a lightcone OPE limit.  The usual conformal cross-ratios are expressed as $u = z \bar z$ and $v = (1-z)(1-\bar z)$. In the $\< \CO_1 \CO_1 \CO_2 \CO_2 \>$ correlator, we refer to conformal blocks exchanged between $\CO_1 \CO_1 \to \CO_2 \CO_2$, such as the $T$ block, as the `t-channel', while we refer to $\CO_1 \CO_2 \to \CO_1 \CO_2$ as the `s-channel', which includes the $[\CO_1 \CO_2]_{n, \ell}$ operators.  As discussed in appendix \ref{sec:2dDecomposition}, we will often make use of a 2d decomposition \cite{KomargodskiZhiboedov} of operators, which makes it easy to handle operators with general spin at minimum twist.

The outline of this paper is as follows. In section \ref{sec:Review} we discuss the setup of the bootstrap equation in the lightcone OPE limit and review previous results on the large $\ell$ spectrum of CFTs. In section \ref{sec:Eikonalization} we show that subleading terms in $1/\ell$ demand the presence of operators like $[TT]_{n, \ell}$ in the $\CO_i(x) \CO_i(0)$ OPE. We then generalize this analysis to demonstrate the `eikonalization' of conformal blocks in this kinematic limit, and use a bootstrap equation to determine the $[TT]_{n, \ell}$ OPE coefficients in section \ref{sec:LargeSpinOPEfromBootstrap}. We then attempt a direct derivation of these OPE coefficients, and identify an obstruction in section \ref{sec:MellinBoundedness}. We show that this problem can be bypassed using Darboux-type arguments, justifying a more direct method for the computation of general large spin OPE coefficients.  We compute general $[TS]_{0, \ell}$ OPE coefficients and use them to verify the leading $\log(\ell)$ behavior of the $[TTT]_{0,\ell}$ coefficients.  We conclude by discussing prospects for future work.

\section{CFT Spectra from the Bottom Up}
\label{sec:Review}

Using our basic `sandbox' of primary operators, we would like to determine as much as possible about the full set of CFT spectra allowed by crossing symmetry and unitarity. In this section, we briefly review previous results that demonstrate our general approach of studying the conformal bootstrap in the lightcone limit, isolating the contributions of large spin operators. 
In this way, we can use a limited collection of operators to discover universal properties of the OPE structure of CFTs. 
Finally, we discuss results from 2d theories where eikonalization can be directly demonstrated using the Virasoro algebra. 

\subsection{Review of Cluster Decomposition and Double-Trace Operators}
\label{sec:ClusterReview}

Let's first consider a general CFT$_d$ containing at least two primary operators, $\Ocal_1$ and $\Ocal_2$. A natural question to ask is whether the presence of these two operators requires additional primary operators in the spectrum. For unitary theories in $d \geq 3$, it was recently proved \cite{AldayMaldacena, Fitzpatrick:2012yx, KomargodskiZhiboedov} that the OPE $\Ocal_1(x) \Ocal_2(0)$ \emph{must} contain an infinite number of large-spin primaries $[\Ocal_1 \Ocal_2]_{n,\ell}$ with scaling dimensions $\De_{n,\ell} \ra \De_1 + \De_2 + 2n + \ell$ as $\ell \ra \infty$.

This tower of integer-spaced scaling dimensions is reminiscent of the spectrum of `double-trace' operators familiar from theories with a perturbative $1/N$ expansion. This recent CFT theorem therefore demonstrates that the Fock space structure of such theories is actually a universal property of CFTs at large spin, consistent with the interpretation of these operators as creating well-separated objects in AdS. Each primary operator then immediately introduces a rich spectrum of large $\ell$ operators to any CFT$_{\geq3}$.

Though this theorem of `cluster decomposition' in CFTs was inspired by the structure of theories in AdS, it can be proven without ever appealing to AdS/CFT. While we will quickly review the basic form of this proof here, interested readers should consult \cite{Fitzpatrick:2012yx, KomargodskiZhiboedov} for a much more detailed discussion.

Any correlation function constructed from the operators $\Ocal_1$ and $\Ocal_2$ can be written as a sum over intermediate states, which can then be organized into irreducible representations of the conformal group as
\be
\<\Ocal_1(\infty) \Ocal_1(1) \Ocal_2(z,\bar{z}) \Ocal_2(0)\> = \fr{1}{(z\bar{z})^{\De_2}} \sum_{\tau,\ell} P^{(11,22)}_{\tau,\ell} \, g_{\tau,\ell}(u,v).
\ee
The individual conformal blocks $g_{\tau,\ell}(u,v)$ are labeled by the spin $\ell$ and twist $\tau \equiv \De - \ell$ of the exchanged primary operators and depend only on the conformally invariant cross-ratios $u=z\bar{z}$ and $v=(1-z)(1-\bar{z})$.
 
As is well-known (e.g. \cite{Rattazzi:2008pe,Fitzpatrick:2014vua}), there are multiple possible channels in which one can decompose a single correlation function into conformal blocks. The equality of these distinct expansions is referred to as the conformal bootstrap equation, which can be written as
\be
u^{-\half(\De_1+\De_2)} \sum_{\tau,\ell} P^{(11,22)}_{\tau,\ell} \, g_{\tau,\ell}(u,v) = v^{-\half(\De_1+\De_2)} u^{-\half \De_{12}} \sum_{\tau,\ell} P^{(12,12)}_{\tau,\ell} \, g_{\tau,\ell}(v,u).
\label{eq:Bootstrap}
\ee
The individual conformal blocks in these series are completely fixed by conformal invariance. More specifically, these blocks can be written as $g_{\tau,\ell}(u,v) = u^{\fr{\tau}{2}} f_{\tau,\ell}(u,v)$, where $f_{\tau,\ell}$ approaches a finite value as $u \ra 0$. We therefore see that in the limit of small $u,v$ these two expansions are dominated by those primary operators with lowest twist.

\begin{figure}[t!]
\begin{center}
\includegraphics[width=0.99\textwidth]{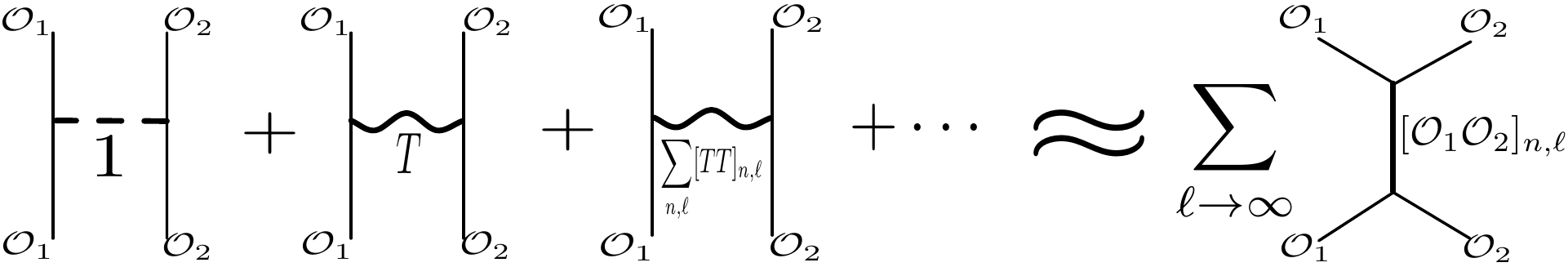}
\caption{ This figure illustrates terms that contribute to the lightcone OPE limit of the CFT bootstrap equation. The consequence of the first two terms on the left-hand side are reviewed in section \ref{sec:Review}, while the third term and its generalizations are discussed using this bootstrap equation in section \ref{sec:Eikonalization}. }
 \label{fig:LightConeLimitBootstrap} 
\end{center}
\end{figure}

Unitarity restricts the possible twists of primary operators, providing the $d$-dependent lower bound
\be
\tau \geq \Bigg\{ \begin{matrix} \fr{d-2}{2} & (\ell = 0), \\ d-2 & (\ell \geq 1). \end{matrix}
\ee
The one exception to this bound is the identity operator, with $\tau = 0$. For $d \geq 3$, unitarity therefore separates the twist of the identity from those of other operators. We can then isolate this universal contribution by taking the limit of small $u$, leading to the approximate relation
\be
u^{-\half(\De_1+\De_2)} \approx v^{-\half(\De_1+\De_2)} u^{-\half \De_{12}} \sum_{\tau,\ell} P^{(12,12)}_{\tau,\ell} \, g_{\tau,\ell}(v,u) \qquad (u \ra 0),
\ee
which is illustrated in figure \ref{fig:LightConeLimitBootstrap}. 
We can clearly see that the left side of the bootstrap equation, which we shall refer to as the `t-channel', possesses a manifest singularity as $u \ra 0$. However, each individual term on the right side, called the `s-channel', is at most logarithmically divergent at small $u$. There must therefore be an infinite number of s-channel conformal blocks, such that the full sum possesses a stronger singularity than any finite combination of terms.

By carefully matching the $u$- and $v$-dependence of both sides, we can then show that both the scaling dimensions and OPE coefficients of these conformal blocks approach those of a generalized free theory as $\ell \ra \infty$. Unitary CFTs therefore possess a universal `weakly-coupled' regime at large $\ell$.

\subsection{Anomalous Dimensions from Minimal Twist Operators}

Though the identity operator provides the dominant t-channel contribution at small $u$, there are corrections from those operators with minimal nonzero twist. For the sake of simplicity, we shall assume that there is only one such operator, though this discussion can easily be generalized to any finite number of minimal twist operators.

Including the correction from the lowest twist operator $T$, the bootstrap equation at small $u$ now takes the approximate form
\be
u^{-\half(\De_1+\De_2)} \bigg( 1 + P^{(11,22)}_T \, g_T(u,v) \bigg) \approx v^{-\half(\De_1+\De_2)} u^{-\half \De_{12}} \sum_{\tau,\ell} P^{(12,12)}_{\tau,\ell} \, g_{\tau,\ell}(v,u).
\label{eq:BSsmallu}
\ee
This additional t-channel block greatly simplifies if we also take the limit $v \ra 0$,
\be
g_T(u,v) \approx - u^{\fr{\tau_T}{2}} \fr{\G(\tau_T+2\ell_T)}{\G^2(\fr{\tau_T}{2}+\ell_T)} \log v \qquad (u \ll v \ll 1),
\ee
where $\tau_T$ and $\ell_T$ are respectively the twist and spin of the minimal twist operator $T$. This conformal block therefore introduces a logarithmic singularity at small $v$ which must be replicated by the s-channel.

To see how this singularity is reproduced, note that the s-channel conformal blocks can be written as $g_{\tau,\ell}(v,u) = v^{\frac{\tau}{2}} f_{\tau,\ell} (v,u)$, where $f_{\tau,\ell}$ is finite as $v \ra 0$. At large $\ell$, we know that the spectrum of s-channel blocks approaches that of the double-trace operators $[\Ocal_1 \Ocal_2]_{n,\ell}$, with the associated twists
\be
\tau(n,\ell) = \De_1 + \De_2 + 2n + \g(n,\ell),
\ee
where the anomalous dimensions $\g(n,\ell) \ra 0$ as $\ell \ra \infty$.

Given this asymptotic behavior, at large $\ell$ we can expand the s-channel conformal blocks as a power series in $\g(n,\ell)$, obtaining
\be
g_{\tau,\ell}(v,u) \approx \left( 1 + \fr{\g(n,\ell)}{2} \log v \right) v^{\fr{\tau_n}{2}} f_{\tau_n,\ell} (v,u) \qquad (\ell \gg 1).
\label{eq:BlockExpansion}
\ee
The anomalous dimensions of double-trace operators therefore provide the logarithmic singularities necessary to match the small $v$ behavior of minimal twist conformal blocks.

As discussed more thoroughly in \cite{Fitzpatrick:2012yx, Fitzpatrick:2014vua, Kaviraj:2015cxa}, we can carefully match the $u$- and $v$-dependence of both sides to precisely fix the anomalous dimensions at large $\ell$. For example, the resulting anomalous dimensions for $n=0$ are 
\be
\label{eq:DefinesGamma0}
\g(0,\ell) \approx - \fr{2 P^{(11,22)}_T \, \G(\De_1) \G(\De_2)\G(\tau_T + 2\ell_T)}{\G(\De_1 - \fr{\tau_T}{2})\G(\De_2 - \fr{\tau_T}{2})\G^2(\fr{\tau_T}{2} + \ell_T)} \left( \fr{1}{\ell^{\tau_T}} \right).
\label{eq:AnomalousDimension}
\ee
The anomalous dimensions therefore vanish as $\ell \ra \infty$ at a rate set by the twist of the exchanged operator $T$. This behavior is consistent with the AdS interpretation of these anomalous dimensions as interaction energies between distant objects.

We also see more explicitly the sense in which CFTs are weakly-coupled at large $\ell$. The corrections to scaling dimensions and OPE coefficients which arise from the exchange of `light' operators with low twist must vanish as $\ell \ra \infty$, therefore introducing the new perturbative parameter $1/\ell$. By continuing the expansion of eq.~(\ref{eq:BlockExpansion}) to higher orders, we can then begin to study subleading corrections due to operators with larger twist.

\subsection{Semi-Classical Virasoro Blocks and Eikonalization}
\label{sec:VirasoroEikonalization}

Our discussion so far has been limited to theories in $d\geq3$. Turning to $d=2$, we see that the unitarity bound no longer separates the identity operator from other minimal twist operators. In the limit $u \ra 0$, we must therefore consider the contribution of not just the identity, but every operator with $\tau = 0$.

One obvious example of such operators is the stress-energy tensor $T_{\mu\nu}$, which provides a universal contribution to our original correlation function in every CFT$_2$. In fact, there is an infinite collection of multi-trace operators constructed out of $T_{\mu \nu}$ which have zero twist and must therefore be included.

However, this significant increase in the number of minimal twist operators is compensated for by the infinite-dimensional Virasoro symmetry of 2d CFTs. All multi-trace operators built from $T_{\mu\nu}$ are Virasoro descendants of the identity and therefore have fixed OPE coefficients. The contribution of these individual global conformal blocks can then be arranged into irreducible representations of the full Virasoro conformal symmetry, referred to as Virasoro blocks.

Even though the structure of Virasoro blocks is entirely fixed by symmetry, they currently have no simple closed form expression that allows one to study their behavior at the edge $z \sim 1 $ of the radius of OPE convergence analytically 
(but see \cite{Perlmutter:2015iya}). However, these blocks greatly simplify in the semi-classical limit of large central charge $c$. In \cite{Fitzpatrick:2014vua}, the general structure of semi-classical Virasoro blocks was studied in the specific limit of $c \ra \infty$ with arbitrary $\De_1$ and fixed $\De_2$. Though the focus of that work was the case $\De_1 \sim c$, so as to study gravitational phenomena associated with black holes in AdS$_3$, it was also shown that for $\De_1,\De_2 \ll c$ the identity Virasoro block takes the simple form
\be
\CV(u,v) \approx \exp \Big[ P^{(11,22)}_T \, g_T(u,v) \Big].
\label{eq:VirasoroExponentiate}
\ee
In other words, the contribution of all multi-trace $T_{\mu\nu}$ operators exponentiates! This form is consistent with the eikonalization of AdS gravitational interactions in the small $G$ limit.  We review and generalize these results, including the effects of graviton interactions, in appendix \ref{sec:2dResults}. We also comment on some interesting cancellations that occur in these calculations, which we have seen persist to order $1/c^2$ at large central charge, involving mixing between one, two, and three `graviton' states.

Note that this eikonalization crucially depends on the `weak-coupling' limit of $c \ra \infty$ and therefore receives corrections suppressed by $1/c$. Regardless, an obvious question is whether this behavior is universal, such that the exchange of other weakly-coupled primary operators also exponentiates.

\section{Eikonalization at Large Spin}
\label{sec:Eikonalization}

Motivated by the semi-classical results in 2d, we now consider multi-trace operators in more general CFTs, in order to determine the generality of eikonalization. As we have seen, the large $\ell$ spectrum of primary operators provides a universal perturbative regime in CFTs, so we expect the OPE coefficients of large-$\ell$ multi-trace operators to be computable.  We begin by considering the subleading corrections due to double-trace primaries constructed from minimal twist operators. We then generalize to the full case of all large $\ell$ multi-trace operators, whose contributions can be resummed as $u \ra 0$. Finally, we briefly compare our CFT results to expectations from the eikonal limit of scattering in AdS.

\subsection{Corrections Due to Large Spin Operators $[TT\cdots T]_{\ell}$}

Let's again consider the anomalous dimensions of the large spin double-trace operators $[\Ocal_1 \Ocal_2]_{n,\ell}$, which arise due to the minimal twist operators present in the OPE of both $\Ocal_1(x) \Ocal_1(0)$ and $\Ocal_2(x)\Ocal_2(0)$. The resulting shift in scaling dimension can then be written in the approximate form
\be
\g(n,\ell) \approx \fr{\g_n}{\ell^{\tau_T}},
\ee
where $\tau_T$ is the minimal twist of these exchanged operators. Using this form, it is clear that at large $\ell$ we can expand the $[\Ocal_1 \Ocal_2]_{n,\ell}$ conformal blocks as a perturbative series in $\g(n,\ell)$,
\be
g_{\tau_n + \gamma(n,\ell),\ell}(v,u) &=&  v^{\half(\tau_n + \gamma(n,\ell))} f_{\tau_n+\gamma(n,\ell),\ell} (v,u) \nn\\
 & \approx & \left( 1 + \fr{\g_n}{2\ell^{\tau_T}} \log v + \fr{\g_n^2}{8\ell^{2\tau_T}} \log^2 v + \cdots \right) g_{\tau_n,\ell}(v,u) \qquad (\ell \gg 1),
 \label{eq:largespinDTblocks}
\ee
where in addition to expanding at small $\gamma_n$, we have kept only the leading log-enhanced terms in  $\gamma_n \log v$.  
As discussed above, the first term in this series reproduces the t-channel contribution from the identity, while the second term contains a logarithmic singularity at small $v$ which matches that of the minimal twist conformal blocks. For the sake of simplicity, we shall again assume that there is only one such minimal twist operator $T$.

Turning to the third term in this series, we see that it possesses a stronger singularity as $v \ra 0$ than any single t-channel conformal block. There must therefore be an infinite tower of additional operators to replicate this subleading correction.  This was already suggested in figure \ref{fig:LightConeLimitBootstrap}.

To determine the properties of this infinite set of t-channel conformal blocks, we first need to determine the full form of the leading-log (LL) s-channel corrections by putting the the large spin double-trace blocks (\ref{eq:largespinDTblocks}) back into the RHS of the bootstrap equation (\ref{eq:BSsmallu}) at small $u$ and summing.  The explicit powers of $v^n$ in $g_{\tau_n + \gamma(n,\ell), \ell}(v,u) \propto v^{n + \half(\Delta_1 + \Delta_2 + \gamma(n,\ell))}$ implies that the dominant contribution to this sum is from the double-trace conformal blocks with $n=0$:
\be
&& v^{-\half(\Delta_1+\Delta_2)} u^{-\half\Delta_{12}} \sum_\ell P^{(12,12)}_{[\Ocal_1 \Ocal_2]_{0,\ell}} g_{\tau_0 + \gamma(0,\ell),\ell}(v,u)  \nn\\
&& \qquad \qquad \stackrel{\rm LL}{=} \sum_{m=0}^\infty u^{-\half(\Delta_1+\Delta_2 - m \tau_T)} \frac{\gamma_0^m \Gamma(\Delta_1 - \frac{m \tau_T}{2}) \Gamma(\Delta_2 - \frac{m\tau_T}{2})}{2^m m! \Gamma(\Delta_1) \Gamma(\Delta_2)} \log^m v ,
\label{eq:LLBS}
\ee
where we have used asymptotic forms of the OPE coefficients and the blocks, and approximated the sum over $\ell$ as an integral.\footnote{Explicit expressions for the asymptotic forms of the OPE coefficients and the blocks in the lightcone limit $u \ll v \ll 1$ at large $\ell$ can be found in \cite{Fitzpatrick:2012yx, Fitzpatrick:2014vua}, which we reproduce here:
\be
g_{\tau,\ell}(v,u) &\approx & v^{\fr{\tau}{2}} u^{\half \De_{12}} 2^{\tau+2\ell} \sqrt{\fr{\ell}{\pi}}  K_{\De_{12}} (2\ell \sqrt{u}) \qquad (u \ll v \ll 1), \nn\\
P^{(12,12)}_{[\Ocal_1 \Ocal_2]_{n,\ell}} &\approx& \fr{4 \sqrt{\pi}}{\G(\De_1) \G(\De_2) 2^{\tau_n+2\ell}} \ell^{\De_1+\De_2-\fr{3}{2}},
\ee
where $K_x(y)$ is a modified Bessel function.
}
Note that the divergent sums over $\ell$ produce very specific singularities as $u \ra 0$. By comparing this $u$-dependence to that of the t-channel, we can then fix the twist of the operators that can reproduce the $m$-th term as $\tau= m \tau_T$.  In particular, we recognize the $m=0$ and $m=1$ terms in parentheses as the contributions due to the identity and $T$ from the t-channel, and we see that the subleading $\log^m v$ corrections must come from infinite towers of operators with $\tau \rightarrow m \tau_T$ as $\ell\rightarrow \infty.$ This is a strong indication that the subleading $\log^2v$ corrections from anomalous dimensions correspond to the large $\ell$ double-trace operators $[T T]_{n,\ell}$!\footnote{At this stage, one might wonder if the role of these infinite towers could be played by some other operators with $\tau = m \tau_T$ besides the $[T T]_{n,\ell}$ double-trace operators.  We will make the connection to the $[T T]_{n, \ell}$ more explicit in sections \ref{sec:LargeSpinOPEfromBootstrap} and \ref{sec:MellinBoundedness}.}


We can then determine the conformal block coefficients for these large $\ell$ operators by matching the $v$-dependence of both sides. The dominant contribution is again from the $n=0$ conformal blocks, with the resulting coefficients
\be
P^{(11,22)}_{[TT]_{0,\ell}} \approx \fr{\g_0^2 \sqrt{\pi}}{2^{2\tau_T+2\ell}} \, \fr{\G(\De_1-\tau_T) \G(\De_2-\tau_T)}{\G(\De_1) \G(\De_2)} \, \ell^{-\fr{3}{2}}.
\ee
Experts may be interested to note that when we multiply this coefficient by the corresponding conformal blocks and perform the sum over $\ell$, the double logarithm $\log^2 v$ arises mainly from the sum over the region $1 \ll \ell^2 \ll \frac{1}{u}$.  

There are many primary operators with large spin $\ell$ and twist near $m \tau_T$ for $m > 2$.  We cannot distinguish among them; we can only obtain the sum of their conformal block coefficients 
at large $\ell$ from our bootstrap argument.  In fact, one can write a general formula for the combined conformal block coefficients of all operators of the form $[T^m]_\ell$ with $m \geq 2$:
\be
\label{eq:FockSpaceBlockCoefficient}
P^{(11,22)}_{\sum [T^m]_{\ell}} \approx \fr{(-\g_0)^m \sqrt{\pi}}{2^{m \tau_T+2\ell} (m-2)! } \, \fr{\G(\De_1- \frac{m\tau_T}{2}) \G(\De_2-\frac{m\tau_T}{2})}{\G(\De_1) \G(\De_2)} \, \ell^{-\fr{3}{2}} \log^{m-2} \left( \ell \right).
\ee
This provides the leading logarithmic dependence of the summed conformal block coefficients at large $\ell$.  We wrote the coefficient in terms of $(-\gamma_0)$ because this quantity is positive, ensuring the manifest positivity of the large spin conformal block coefficients when $\CO_1 = \CO_2$.  We will obtain the $m=3$ case directly from an OPE limit in section \ref{sec:UniversalityandDarboux}.

\subsection{Exponentiation of Large Spin Operators and AdS Field Theory}


Next, let us consider the $m>2$ terms in (\ref{eq:LLBS}).  Since these correspond to infinite towers of operators with twist $\tau \rightarrow m \tau_T$ as $\ell\rightarrow \infty$, they are most naturally interpreted as multi-trace operators constructed from $T$.  
%
We therefore find that in \emph{every} CFT$_{\geq3}$, if a single minimal twist operator $T$ contributes to the correlation function $\<\Ocal_1 \Ocal_1 \Ocal_2 \Ocal_2\>$, then there must also be a universal multi-trace contribution of the form
\be
\sum_{m=0}^\infty P^{(11,22)}_{[T^m]_\ell} \, g_{[T^m]_\ell}(u,v) \approx \sum_{m=0}^\infty \fr{1}{m!} \left( \fr{\g_0}{2} u^{\fr{\tau_T}{2}} \log v \right)^m \fr{\G(\De_1-m\fr{\tau_T}{2}) \G(\De_2-m\fr{\tau_T}{2})}{\G(\De_1) \G(\De_2)},
\ee
where we have specifically taken the limit $u \ll v \ll 1$.

This expression greatly simplifies if we consider the limit $\De_1,\De_2 \gg \tau_T$, such that two `heavy' operators are exchanging a `light' mediator. In this case, we obtain
\bq
\begin{split}
\sum_{m=0}^\infty P^{(11,22)}_{[T^m]_\ell} \, g_{[T^m]_\ell}(u,v) &\approx \sum_{m=0}^\infty \fr{1}{m!} \left( -P^{(11,22)}_T \, u^{\tau_T} \fr{\G(\tau_T+2\ell_T)}{\G^2(\fr{\tau_T}{2}+\ell_T)} \log v \right)^m \\
&\approx \exp \Big[ P^{(11,22)}_T \, g_T(u,v) \Big],
\end{split}
\label{eq:expeik}
\eq
where we have used eq.~(\ref{eq:AnomalousDimension}) to rewrite the anomalous dimension coefficient $\g_0$. 
Note that at small $u, v$ this behaves as $v^{-\alpha' u^{\tau_T}}$ with $\alpha' \equiv P_T^{(11,22)} \frac{\Gamma(\tau_T+2\ell_T)}{\Gamma^2(\frac{\tau_T}{2}+\ell_T)}$.

We see that the exchange of a minimal twist operator exponentiates in the lightcone OPE limit. This universal eikonalization arises due to the fact that the kinematic limit $v \ra 0$ isolates the large $\ell$ multi-trace operators, which form an approximate Fock space. The exchange of these multi-trace operators in a CFT is therefore consistent with the exchange of well-separated, weakly-interacting light mediators in AdS.  These exchanges are what is responsible for the fact that one can replace the `source' with its effective classical background field (see e.g. chapter 13 of \cite{Weinberg:1995mt}).  
A similar analysis can also be perfomed in the usual eikonal limit in AdS \cite{CostaEikonal}.  In that case it can be shown that AdS/CFT correlators can be written as a bulk integral over the exponential of an effective eikonal propagator \cite{CostaEikonal} in a $d-1$ dimensional hyperbolic space.  Although we have not been studying the eikonal limit of large energy and large impact parameter, but instead have focused on fixed energy and large impact parameter, it would be interesting to connect to these AdS results more concretely in future work. 

\subsection{Explicit Construction of Large-$\ell$ `Multi-Trace' Modes}

The result (\ref{eq:expeik}) generalizes those of \cite{Fitzpatrick:2014vua}, which focused on the case of the stress tensor in $d=2$.  An advantage of the latter approach was that all $[T \dots T]_\ell$ contributions could be constructed explicitly using generators of the Virasoro algebra, so that it was manifest which modes were exponentiating.  In this section, we will generalize the construction to an arbitrary spin-$L$ current $J$ in $d=2$.  In this case, the commutators of modes $J_n$ of the current are simple to calculate as a result of holomorphicity.  Conserved currents have zero twist in $d=2$, so according to (\ref{eq:expeik}), their contributions should exponentiate assuming the anomalous dimensions can be treated perturbatively.  

The singular terms in the OPE of a general spin-$L$ current contain at a minimum
\be
J(z) \Ocal_i(w) \sim q_i \sum_{a=0}^{L-1} \frac{1}{(z-w)^{L-a}} \frac{(L)_a}{a!(2h_i)_a} \partial^a \Ocal_i(w),
\ee
where $q_i$ is the charge of $\Ocal_i$. In what follows, for convenience of notation we assume these are the only singular terms.  Writing $J(z) = \sum_{n\in \mathbb{Z}} \frac{J_n}{z^{n+L}}$, we can read off the commutators by a standard contour integral,
\begin{equation}
[J_n, \Ocal_i(w)] = \oint \frac{dz}{z} z^{n+L} J(z) \Ocal_i(w) = q_i \sum_{a=0}^{L-1}  \frac{\Gamma(L+n)(L)_a w^{a+n}}{\Gamma(L-a)(n+a)! a! (2h_i)_a} \partial^a \Ocal_i(w).
\end{equation}
The contribution of the modes $J_n$ to the four-point function $\< \Ocal_1^\dagger \Ocal_1 \Ocal_2 \Ocal_2^\dagger\>$ depends on the matrix of inner products as well as the matrix elements $\< \Ocal_1^\dagger \Ocal_1 J_{-n}\>, \< J_n \Ocal_2 \Ocal_2^\dagger\>$.  The derivatives $\partial^a$ acting on $\Ocal_i(w)$ inside $\< \Ocal_i \Ocal_i^\dagger\>$ produces 
\be
w^{a+n} \partial^a \< \Ocal_i(w) \Ocal_i^\dagger(0)\> \rightarrow w^n (-1)^a (2h_i)_a \< \Ocal_i (w) \Ocal_i^\dagger(0)\> ,
\ee
so we can simplify the action of $J_n$ in such matrix elements to
\be
\< J_n \Ocal_i(w) \Ocal_i^\dagger(0)\> &=& q_i w^n \sum_{a=0}^{L-1} \frac{\Gamma(L+n)(L)_a(-1)^a }{\Gamma(L-a)(n+a)! a! } \< \Ocal_i(w) \Ocal_i^\dagger(0)\> \nn\\
 &=& q_i w^n \left( n-1 \atop L-1 \right)\< \Ocal_i(w) \Ocal_i^\dagger(0)\> .
\ee
The algebra of higher-spin currents in general can be extremely complex; however, here we are interested in the limit where the central charge of the current is large, so the dominant term will be given by the $\<J(z)J(0)\>$ two-point function:
\be
\< J(z) J(0) \> &=& \frac{c_J}{z^{2L}}.
\ee
To leading order, the algebra is therefore approximately
\be
[ J_n, J_m] \approx c_J \frac{\Gamma(L+n)}{\Gamma(2L)\Gamma(n-L+1)} \delta_{n,-m}.
\ee
The contribution from a single exchange immediately reproduces the standard conformal block:
\be
\sum_{n=L}^\infty \frac{\< \Ocal_1^\dagger(\infty) \Ocal_1(1) J_{-n}\>  \< J_n \Ocal_2(z) \Ocal_2^\dagger(0)\> }{\< J_n J_{-n}\>}
 &=& \< \Ocal_1^\dagger(\infty) \Ocal_1(1) \>  \< \Ocal_2(z) \Ocal_2^\dagger(0)\> \frac{q_1q_2}{c_J} z^L {}_2F_1(L,L,2L,z). \nn\\
\ee
In the limit of large $c_J$ with $\frac{q_1 q_2}{c_J}$ fixed, all $J_n$'s with positive $n$ commute with each other at leading order, and similarly for $J_n$'s with negative $n$, so as in \cite{Fitzpatrick:2014vua} the contribution from  
modes made of all products of $J_n$ is just the exponentiation of the single-$J$ contribution:
\be
&& \sum_{s=0}^\infty \sum_{n_1, n_2, \dots n_s=L}^\infty\frac{\< \Ocal_1^\dagger(\infty) \Ocal_1(1) J_{-n_1}\dots J_{-n_s}\>  \< J_{n_s} \dots J_{n_1} \Ocal_2(z) \Ocal_2^\dagger(0)\> }{\< J_{n_s} \dots J_{n_1} J_{-n_1} \dots J_{-n_s}\>} \nn\\
 && = \< \Ocal_1^\dagger(\infty) \Ocal_1(1) \>  \< \Ocal_2(z) \Ocal_2^\dagger(0)\> \exp \left[ \frac{q_1q_2}{c_J} z^L {}_2F_1(L,L,2L,z) \right]. 
\ee

\section{Large Spin OPE Coefficients from a Boostrap Analysis}
\label{sec:LargeSpinOPEfromBootstrap}


The eikonalization of conformal blocks indicates a connection between the $\Ocal_i(x) \Ocal_i(0)$ OPE coefficients for $T$ and its multi-trace counterparts $[T\cdots T]_\ell$. In this section, we make this connection manifest by deriving the large $\ell$ double-trace coefficients from the contribution of $\Ocal_i$ exchange to the correlator $\<\Ocal_i \Ocal_i TT\>$. By studying this single conformal block in the lightcone OPE limit, we successfully reproduce the coefficients derived in section~\ref{sec:Eikonalization}. We then consider the exchange of  additional conformal blocks, demonstrating the effect on eikonalization of finite numbers of such potential corrections.

\subsection{Exchange of $\Ocal_i$ in the Lightcone OPE Limit}
\label{sec:TTfromLightcone}


So far, we have obtained an indirect relation between the OPE coefficients of the single-trace operator $T$ and the multi-trace $[T \cdots T]_\ell$ by using the bootstrap equation for the $\< \CO_1 \CO_1 \CO_2 \CO_2 \>$. However, as indicated in figure~\ref{fig:FromBlocksToOPE}, we can instead derive this connection by considering correlation functions involving multiple insertions of $T$. 

For example, given a four-point function of the form $\<\Ocal_i \Ocal_i TT\>$, we can take the kinematic limit $z\ra1$, $\bar{z}\ra0$ to determine the OPE coefficients for all of the double-trace operators $[TT]_{n,\ell}$. While the precise form of this correlation function clearly depends on the dynamics of the particular theory being studied, we are specifically interested in contributions associated with the presence of $T$ in the OPE $\Ocal_i(x) \Ocal_i(0)$.

One universal such contribution is the exchange of the $\Ocal_i$ conformal block, shown schematically in figure \ref{fig:TwoNecessaryBlocks}, which is completely fixed by the corresponding OPE coefficient for $T$. A natural question to ask is whether this conformal block gives rise to the universal double-trace coefficients derived in the previous section. While we will specifically consider this calculation for $d=2$, as this is the simplest case technically, the results can easily be generalized to higher dimensions.

Our strategy will therefore be quite analogous to our method in section \ref{sec:Eikonalization}.  In the limit $z\ra 1$, the LHS of our bootstrap equation is dominated by those operators with lowest holomorphic dimension $h$.  We begin with singularities that arise from the exchange of a single operator (or more generally later on, a finite number of operators), on one side of the bootstrap equation, and ask what must appear on the other side in order to reproduce it. For now, we will simply assume that this operator is $\Ocal_i$, but in the following section we'll consider corrections due to the presence of additional operators $\Ocal'$ in the OPE $\Ocal_i(x) T(0)$.
   Expanding the correlation function $\<\Ocal_i \Ocal_i TT\>$ in two independent channels, we can then obtain the bootstrap equation in the limit $z \ra 1$:
\be
P^{(\Ocal_i T,\Ocal_i T)}_{\CO_i} \, g_{\CO_i}(1-z,1-\bar{z}) \approx \fr{(1-z)^{h_i+h_T} (1-\bar{z})^{\bar{h}_i+\bar{h}_T}}{z^{2h_T} \bar{z}^{2\bar{h}_T}} \sum_{h,\bar{h}} P^{(\Ocal_i \Ocal_i,TT)}_{h,\bar{h}} \, g_{h,\bar{h}}(z,\bar{z}), \quad
\label{eq:OOTTBootstrap}
\ee
which is much like eq. (\ref{eq:BSsmallu}), except that now $\CO_i$ and $T$ are external operators and $\CO_i$ is the internal operator.  Just as in the derivation of cluster decomposition reviewed in section~\ref{sec:Review}, the LHS cannot be reproduced at $z\sim 1$ by any finite number of conformal blocks on the RHS, implying the presence of an infinite number of conformal blocks in the cross-channel.

To make this more explicit, we again take limits.  At $\bar{z} \rightarrow 0$ and $h\rightarrow \infty$ with fixed $h\sqrt{1-z}$, eq. (\ref{eq:OOTTBootstrap}) approximates to\footnote{
At $\bar{z} \ra 0$, the global conformal block associated with the exchange $\CO_i T \to \Ocal_i \to \CO_i T$ takes the approximate form
\be \label{eq:LightConeforTT}
g_{\Ocal_i}(1-z,1-\bar{z}) \approx \fr{\G(2\bar{h}_i) \G(2\bar{h}_i-2\bar{h}_T)}{\G^2(2\bar{h}_i-\bar{h}_T)} \, (1-z)^{h_i} \qquad (1-z \ll \bar{z} \ll 1),
\ee
and the approximate conformal blocks for the large $h$ contribution on the RHS are
\be
\label{eq:LargehBlocks}
g_{h,\bar{h}}(z,\bar{z}) \approx \bar{z}^{\bar{h}} \fr{\G(2h)}{\G^2(h)} \, K_0 \left( 2h\sqrt{1-z} \right) + \CO(\bar{z}^{\bar{h}+1}) \qquad (h \gg 1, \ h \sqrt{1-z} \textrm{ fixed} ).
\ee
}
\begin{equation}
P^{(\Ocal_i T,\Ocal_i T)}_{\Ocal_i} \, \fr{\G(2\bar{h}_i) \G(2\bar{h}_i-2\bar{h}_T)}{\G^2(2\bar{h}_i-\bar{h}_T)} \approx \frac{(1-z)^{h_T}}{\bar{z}^{2\bar{h}_T}}  \sum_{h,\bar{h}} P^{(\Ocal_i \Ocal_i,TT)}_{h,\bar{h}}\left( \bar{z}^{\bar{h}} 2^{2h-1} \sqrt{\frac{h}{\pi}} K_0(2h \sqrt{1-z}) + \CO(\bar{z}^{\bar{h}+1})\right). 
\end{equation}
Note that the $\bar{z}$-dependence has greatly simplified, such that we can determine the antiholomorphic dimension $\bar{h}$ of the operators which dominate at large $h$. More precisely, there must be an infinite tower of operators with $\bar{h} \ra 2 \bar{h}_T$ as $h \ra \infty$, in order to cancel the $\bar{z}$-dependent factor.

By matching the full $\bar{z}$-dependence of both sides of the bootstrap equation, we can in fact prove that there must be an infinite tower of such operators for every non-negative integer $n$, with the corresponding dimensions $\bar{h} \ra 2\bar{h}_T + n$. These towers of operators precisely correspond to the large spin double-trace operators $[TT]_{n,\ell}$, with 
\be
h \ra 2h_T + \ell, \qquad \bar{h} \ra 2 \bar{h}_T+n.
\ee
We can then use the $z$-dependence of this infinite sum to determine the asymptotic form of their conformal block coefficients.

As a simple example, let's consider the lowest-twist operators, with $n=0$. 
Parametrizing the large spin conformal block coefficients as
\be
P^{(\Ocal_i \Ocal_i,TT)}_{[TT]_{0,\ell}} \approx \fr{P_0}{2^{4h_T+2\ell}} \, \ell^\alpha,
\ee
we can approximate the sum over $\ell$ as an integral, obtaining
\be
\label{eq:IntegralOverBessel}
\fr{P_0}{2\sqrt{\pi}} \int d\ell \, \ell^{\alpha+\half} K_0(2\ell\sqrt{1-z}) \approx \fr{P_0 \,\G^2(\fr{\alpha}{2}+\fr{3}{4})}{8\sqrt{\pi}} (1-z)^{-\fr{\alpha}{2}-\fr{3}{4}}.
\ee

The divergent sum over large $\ell$ operators therefore leads to a singularity in $z$ not possessed by any finite collection of conformal blocks. By matching the $z$-dependence of the bootstrap equation, we then obtain the large $\ell$ conformal block coefficients
\be
P^{(\Ocal_i \Ocal_i,TT)}_{[TT]_{0,\ell}} \approx \fr{8\sqrt{\pi} \, P^{(\Ocal_i T,\Ocal_i T)}_{\Ocal_i} \, \G(2\bar{h}_i) \G(2\bar{h}_i-2\bar{h}_T)}{\G^2(h_T) \G^2(2\bar{h}_i-\bar{h}_T) 2^{4h_T+2\ell}} \, \ell^{2h_T-\fr{3}{2}}.
\ee
We therefore see that the $[TT]_{n,\ell}$ coefficients are fixed in terms of the $\Ocal_i$ conformal block coefficient.

However, we still need to determine whether these conformal block coefficients are consistent with the eikonal results derived in section~\ref{sec:Eikonalization}. In order to compare conformal blocks associated with different correlation functions, we need to first rewrite them in terms of the more universal OPE coefficients. For example, these particular coefficients can be written as the product
\be
P^{(\Ocal_i \Ocal_i,TT)}_{[TT]_{0,\ell}} = C^{(\Ocal_i \Ocal_i)}_{[TT]_{0,\ell}} C^{(TT)}_{[TT]_{0,\ell}}.
\ee
We can therefore use the known OPE coefficients \cite{Unitarity}
\be
C^{(TT)}_{[TT]_{0,\ell}} \approx \fr{2 (4\pi)^{\fr{1}{4}}}{\G(2h_T) 2^{2h_T+\ell}} \, \ell^{2h_T-\frac{3}{4}},
\ee
to obtain
\be
\label{eq:TTOPECoeff}
C^{(\Ocal_i \Ocal_i)}_{[TT]_{0,\ell}} \approx \fr{2(4\pi)^{\fr{1}{4}} \, P^{(\Ocal_i T,\Ocal_i T)}_{\Ocal_i} \, \G(2h_T) \G(2\bar{h}_i) \G(2\bar{h}_i-2\bar{h}_T)}{\G^2(h_T) \G^2(2\bar{h}_i-\bar{h}_T) 2^{2h_T+\ell}} \, \ell^{-\fr{3}{4}}.
\ee

In section~\ref{sec:Eikonalization}, we considered the contribution of these same large spin operators to the 4-pt correlator $\<\Ocal_1 \Ocal_1 \Ocal_2 \Ocal_2\>$. As a reminder, those conformal block coefficients were found to be
\be
P^{(11,22)}_{[TT]_{0,\ell}} \approx \fr{8\sqrt{\pi} \left( P^{(11,22)}_T \right)^2 \G^2(2h_T)\G(2\bar{h}_1)\G(2\bar{h}_2)}{\G^4(2h_T) \G(2\bar{h}_1-2\bar{h}_T) \G(2\bar{h}_2-2\bar{h}_T)2^{4h_T+2\ell}} \, \ell^{-\fr{3}{2}}.
\ee
By noting the equivalence
\be
\left( P^{(11,22)}_T \right)^2 = \left( C^{(11)}_T C^{(22)}_T \right)^2 = P^{(1T,1T)}_1 P^{(2T,2T)}_2,
\ee
we can then confirm that our OPE coefficients precisely agree with the conformal block coefficients derived in section \ref{sec:Eikonalization}, with the relation
\be
P^{(11,22)}_{[TT]_{0,\ell}} = C^{(11)}_{[TT]_{0,\ell}} C^{(22)}_{[TT]_{0,\ell}} \sim \left( C^{(11)}_T C^{(22)}_T \right)^2.
\ee

We have a simple method for deriving the OPE coefficients of multi-trace operators directly from the coefficients of their constituent primary operators. For any operator $T$ present in the OPE $\Ocal_i(x) \Ocal_i(0)$, the OPE coefficients for the double-trace operators $[TT]_{n,\ell}$ can be determined by considering the correlation function $\<\Ocal_i \Ocal_i TT\>$ in the lightcone OPE limit. This method can easily be generalized to study higher-trace operators, with the resulting OPE coefficients leading directly to the eikonalization of multi-trace operators at large spin.

\subsection{Including $\CO'$ in the $\CO_i(x) T(0)$ OPE}

\begin{figure}[t!]
\begin{center}
\includegraphics[width=0.95\textwidth]{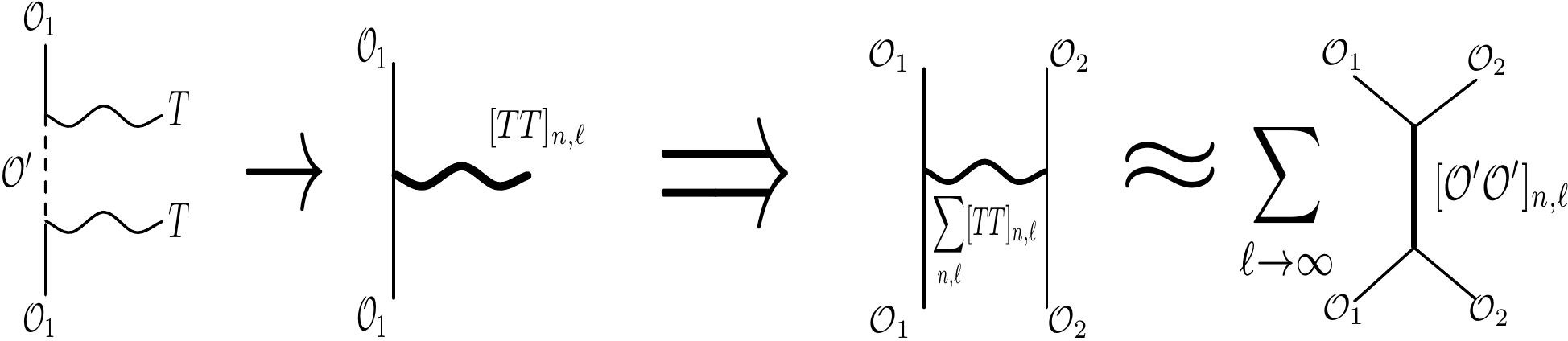}
\caption{ This figure suggests other contributions to the $\< \CO_1 \CO_1 T T \>$ correlator from $\CO'$ conformal blocks.  These contributions affect the $\< \CO_1 \CO_1 [TT]_{n, \ell} \>$ OPE coefficients, but they do not contaminate eikonalization unless $\tau_{\CO'} < \tau_1$ and the two twists differ by an integer.  Generically, they contribute to $[\CO' \CO']_{n, \ell}$ exchange at large $\ell$ in the cross-channel, as pictured, and also to $[\CO_i \CO']_{n,\ell}$. }
 \label{fig:OPrimeContributionTT} 
\end{center}
\end{figure}

Now we will address what would seem to be a major oversight in the previous sections: do other operators $\CO'$ contaminate and invalidate the eikonalization results?

For example, consider the addition of a single new operator $\CO'$ in the OPE $\CO_i(x) T(0)$.  Its presence implies the existence of a conformal block of the form $\CO_i T \to \CO' \to \CO_i T$ contributing to the $\< \CO_i \CO_i T T \>$ correlator, as pictured in figure \ref{fig:OPrimeContributionTT}.  We are particularly interested in how $\CO'$ affects the OPE coefficients of the double-trace operators $[TT]_{n,\ell}$.  We can study the contribution from $\CO'$ by generalizing the bootstrap analysis of the previous section, comparing $\Ocal'$ exchange in one channel to $[TT]_{n,\ell}$ exchange in the cross-channel.

Including the contribution of a finite number of operators like $\CO'$, we can write a bootstrap equation similar to eq.~(\ref{eq:OOTTBootstrap}),
\be
\sum_{\CO'} P^{(\Ocal_i T,\Ocal_i T)}_{\CO'} \, g_{\CO'}(1-z,1-\bar{z}) = \fr{(1-z)^{h_i+h_T} (1-\bar{z})^{\bar{h}_i+\bar{h}_T}}{z^{2h_T} \bar{z}^{2\bar{h}_T}} \sum_{h,\bar{h}} P^{(\Ocal_i \Ocal_i,TT)}_{h,\bar{h}} \, g_{h,\bar{h}}(z,\bar{z}). \qquad
\ee
In the lightcone OPE limit $z \to 1, \bar z \to 0$, the individual terms on the LHS have the leading dependence
\be
g_{\Ocal'} (1-z,1-\bar{z}) \approx \fr{\G(2\bar{h}_i) \G(2\bar{h}_i-2\bar{h}_T)}{\G^2(\bar{h}_i +  \bar{h}_{\Ocal'} -\bar{h}_T)} \, (1-z)^{h_{\Ocal'}} \qquad (1-z \ll \bar{z} \ll 1),
\ee
where the sub-leading terms have a series expansion in $\bar z$ and $1-z$.  
Using an analysis identical to that of the previous section, we can derive a contribution $\delta C$ to the $\Ocal_i(x) \Ocal_i(0)$ OPE coefficients of $[TT]_{0,\ell}$ of the form  
\be
\label{eq:}
\delta C^{(\Ocal_i \Ocal_i)}_{[TT]_{0,\ell}} \approx \fr{2(4\pi)^{\fr{1}{4}} \, P^{(\Ocal_i T,\Ocal_i T)}_{\Ocal'} \, \G(2h_T) \G(2\bar{h}_i) \G(2\bar{h}_i-2\bar{h}_T)}{\G^2(h_T + h_i - h_{\CO'}) \G^2(\bar{h}_i +  \bar{h}_{\Ocal'} -\bar{h}_T) 2^{2h_T+\ell}} \, \ell^{h_i - h_{\CO'} - \fr{3}{4}}.
\ee
These results demonstrate that as $z \to 1$, or equivalently as $\ell \to \infty$, the dominant $\CO'$ will have minimum $h_{\CO'}$.  In particular, any operator $\CO'$ with $h_{\CO'} < h_i$ will dominate over the eikonal contribution to the $[TT]_{n,\ell}$ OPE coefficients from $\CO_i$ itself.  

This does not invalidate the results of the previous sections, but it shows that there are other important contributions to the OPE coefficients $ C^{(\Ocal_i \Ocal_i)}_{[TT]_{n,\ell}}$. Both the contributions discussed in section \ref{sec:TTfromLightcone} and here will be positive, since they are proportional to the square of $\Ocal_i(x) T(0)$ OPE coefficients. The sum of $[TT]_{n,\ell}$ conformal blocks therefore contains two physically distinct pieces. Only the contribution due to $\Ocal_i$ exchange matches the exponentiation of anomalous dimensions discussed in section \ref{sec:Eikonalization}.  As we will see, the $\CO'$ contributions can be matched to other effects.  We will demonstrate how this works for an individual $\CO'$, and the argument can be generalized  as well to any finite number of $\CO'$. We expect the methods of the section \ref{sec:UniversalityandDarboux} to apply even to infinite towers of $\CO'$ with positive twist, although there may be subtleties in that case that deserve a more thorough study.  
   
Let us now argue that, as pictured on the right in figure \ref{fig:OPrimeContributionTT}, the $\CO'$ contributions to the OPE coefficients $C^{\Ocal_i\Ocal_i}_{[TT]_{n,\ell}}$ can be matched by the exchange of additional double-trace operators $[\CO' \CO_i]_{n,\ell}$ and $[ \CO' \CO' ]_{n\ell}$ in the bootstrap. Returning to the correlation function $\<\Ocal_1\Ocal_1\Ocal_2\Ocal_2\>$, the conformal block coefficients associated with $[TT]_{n,\ell}$ exchange are now
\be
P^{(11,22)}_{[TT]_{n,\ell}} = \left( C^{(11)}_{[TT]_{n,\ell}} + \delta C^{(11)}_{[TT]_{n,\ell}} \right) \left( C^{(22)}_{[TT]_{n,\ell}} + \delta C^{(22)}_{[TT]_{n,\ell}} \right),
\ee
so there are four terms that we can study. The original $C \times C$ term simply corresponds to the eikonal results of section~\ref{sec:Eikonalization}, reproducing the exponentiated anomalous dimension in the cross-channel. Turning to the $\delta C \times \delta C$ term, specifically for the minimal twist operators with $n=0$, we find that it contributes
\be
\sum_\ell P^{(11,22)}_{[TT]_{0,\ell}} g_{[TT]_{0,\ell}} (z, \bar z) \supset \sum_\ell \frac{ \ell^{h_1 + h_2 - 2 h_{\CO'} - \frac{3}{2} } }{2^{4 h_T + 2 \ell}} g_{[TT]_{0,\ell}} (z, \bar z),
\ee
where we have neglected to display the overall constant coefficient for simplicity.  We can approximate the sum using equations (\ref{eq:LargehBlocks}) and (\ref{eq:IntegralOverBessel}) with $h = 2 h_T + \ell$ and $\bar h = 2 \bar h_T$.  In the limit that $\bar z$, $1-z$ are small, the result is\footnote{This result holds in the limit $h_1 + h_2 - 2 h_{\CO'} > 0$, where the integral is dominated by large $\ell$ as $z \to 1$.}
\be
\frac{\bar z^{2 \bar h_T} }{2 \sqrt \pi} \int d\ell \, \ell^{h_1 + h_2 - 2 h_{\CO'} - 1} K_0(2\ell\sqrt{1-z}) \approx 
\frac{ \G^2(\fr{h_1 + h_2}{2} - h_{\CO'})}{8 \sqrt{\pi}}  \fr{\bar z^{2 \bar h_T}}{(1-z)^{\fr{h_1 + h_2}{2} - h_{\CO'} }} .
\ee
We have obtained a new singularity structure in the limit $z \to 1$, which is the lightcone OPE limit of $\CO_1(1)\CO_2(z)$.  The power-law of this singularity encodes the presence of the new double-trace operators $[\CO' \CO']_{n,\ell}$ in the OPE.

Similarly, if we study the contribution of the $C^{(11)} \times \delta C^{(22)}$ cross term then we obtain a power-law growth of $\ell^{h_2 - h_{\CO'}}$. This growth at large $\ell$ introduces a new lightcone singularity of $(1-z)^{\fr{h_2 - h_{\CO'}}{2} }$, corresponding to the contribution of $[\CO_1 \CO']_{n,\ell}$ in the $\CO_1(1) \CO_2(z)$ lightcone OPE. The final $C^{(22)} \times \delta C^{(11)}$ term then leads to the contribution of $[\CO_2 \CO']_{n,\ell}$. The only exception to these results is if $h_{\CO'}  = h_1$ or $h_2$, or if they differ by an integer.  

In summary, we can naturally account for the presence of $\CO_i T \to \CO' \to \CO_i T$ contributions to the $[TT]_{n,\ell}$ OPE coefficients. Rather than spoil the eikonalization of $T$ exchange in the t-channel of $\<\Ocal_1\Ocal_1\Ocal_2\Ocal_2\>$, these contributions reproduce the exchange of $[\CO_i \CO']_{n,\ell}$ and $[\CO' \CO']_{n,\ell}$ in the s-channel, as suggested in figure \ref{fig:OPrimeContributionTT}.  As a corollary, we see again that the behavior of the $[TT]_{n,\ell}$ OPE coefficients at large $\ell$ tells us about the dimensions of the large spin operators in the $\CO_1(1) \CO_2(z)$ OPE.  Conversely, knowing the dimensions of operators in the $\CO_1(1) \CO_2(z)$ OPE gives a wealth of information about cross-channel OPE coefficients.

\section{OPE Limits and Direct Extraction of Large Spin Operators} 
\label{sec:MellinBoundedness}



We have demonstrated that the leading large $\ell$ behavior of the OPE coefficients of multi-trace operators such as $[TT \cdots ]_{\ell}$ are fixed by the OPE coefficients of their single-trace constituents (i.e. $T$) through the leading singularities that they imply in the $\CO_i T \rightarrow \CO_i T$ channel.  In this section, we study this connection from a different perspective by considering the $T(x) T(0)$ OPE limit of the correlator $\< \CO_i \CO_i T T\>$.
One might hope that the results in the previous section could be derived more directly, by taking the $T(x) T(0)$  OPE limit  of conformal blocks such as $\CO_i T \rightarrow \CO_i \rightarrow \CO_i T$.   However, we encounter a well-known obstruction in section \ref{sec:OPElimit}, as individual conformal blocks have non-analyticities in the cross-channel OPE limit that are not consistent with unitarity.  
%
%


This non-analyticity has a simple manifestation in Mellin space, as we discuss in section \ref{sec:OPEandMellin}, where we find a direct connection between the cross-channel OPE limit and the asymptotic behavior of the Mellin amplitude.  We use this connection to argue that the $[TT]_{n,\ell}$ OPE coefficients at large $\ell$ will always be determined by the operators in the $\CO_i(x) T(0)$ OPE with twist smaller than $\tau_i + \tau_T$.  In particular, we confirm that there is a universal contribution from the presence of $\CO_i$ itself in this OPE.  Then in section \ref{sec:UniversalityandDarboux} we use Darboux-type arguments to exploit this universality and compute Fock space OPE coefficients using differential operators.

\subsection{Conformal Blocks in Their Cross-Channel OPE Limit}
\label{sec:OPElimit}


In the previous sections, we used the conformal bootstrap to obtain constraints on the OPE coefficients of multi-trace operators $[T \cdots T]_{\ell}$, fixing these contributions in terms of the coefficients of $T$. We can instead try to derive this connection directly, by using the structure of these large $\ell$ operators in the OPE limit.

For simplicity, let us focus on the double-trace operators with minimal twist, i.e. $n=0$. At large $\ell$, the structure of these operators approaches that of a generalized free theory, with the schematic form
\be
[TT]_{0,\ell} \sim T \partial_{\mu_1} \cdots \partial_{\mu_\ell} T,
\ee
where their anomalous dimensions vanish as $\ell \to 0$.
This structure suggests that we can ``build'' double-trace correlation functions out of those containing two insertions of $T$. More concretely, for a given $\ell$ there exists a differential operator $\Dcal_\ell$ such that
\be
\<\Ocal_i(\infty) \Ocal_i(1) | [TT]_{0,\ell}\> \approx \lim_{z,\bar{z} \ra 0} \Dcal_\ell \<\Ocal_i(\infty) \Ocal_i(1) T(z,\bar{z}) T(0)\>,
\ee
with the specific form of $\Dcal_\ell$ fixed by the requirement that $[TT]_{0,\ell}$ is primary.   Note that this procedure can only be applied at large $\ell$.  Only averages over $[TT]_{0,\ell}$ OPE coefficients at large spin are universal, as one can show using the Hardy-Littlewood Tauberian theorem \cite{Fitzpatrick:2012yx, Fitzpatrick:2014vua} via a more rigorous analysis.  

This ``direct'' approach provides an alternative means of deriving the OPE coefficients for large spin $[TT]_{0,\ell}$ from the four-point function $\<\Ocal_i\Ocal_i TT\>$. If the OPE limit $z,\bar{z}\ra0$ of the full correlator is well-defined, the OPE coefficients calculated in this way must agree with those obtained in the previous section from a bootstrap analysis. To confirm this consistency, we need to consider the action of our differential operator $\Dcal_\ell$ on the $\Ocal_i$ conformal block.  

We will confine ourselves to computations of leading twist OPE coefficients, so that we can take advantage of the simplicity of 2d conformal blocks.  We label operators using holomorphic and antiholomorphic dimensions $h,\bar{h}$ and write the leading twist part of the $\Ocal_i$ conformal block as 
\be
g_{\Ocal_i}(z,\bar z) = (1-z)^{h_i} (1-\bar{z})^{\bar{h}_i} \phantom{}_2 F_1 \left( h_T,h_T;2h_i;1-z \right) \, \phantom{}_2 F_1 \left( \bar{h}_T,\bar{h}_T;2\bar{h}_i;1-\bar{z} \right),
\ee
The differential operator $\Dcal_\ell$ can be split into holomorphic and antiholomorphic pieces, such that we have
\be
\Dcal_\ell = \Dcal_{\ell,z} + \Dcal_{\ell,\bar{z}}.
\ee
While the precise form of these differential operators is discussed in section \ref{sec:UniversalityandDarboux}, schematically, they consist of $\ell$ derivatives $\Dcal_{\ell,z} \sim \partial_z^\ell$.
To determine the action of $\Dcal_\ell$, we need to evaluate expressions of the form
\be
\partial_z^\ell \, \left[ (1-z)^{h_i} \phantom{}_2 F_1 \left( h_T,h_T;2h_i; 1-z \right) \right].
\ee
We can simplify this analysis by using the hypergeometric identity
\bq
\begin{split}
\phantom{}_2 F_1 (a,b;c;1-z) &= \fr{\G(c)\G(c-a-b)}{\G(c-a)\G(c-b)} \phantom{}_2 F_1(a,b;1+a+b-c;z) \\
& \, + \, \fr{\G(c) \G(a+b-c)}{\G(a)\G(b)} z^{c-a-b} \phantom{}_2 F_1(c-a,c-b;1+c-a-b;z).
\end{split}
\label{eq:HyperIdentity}
\eq
Acting with derivatives on the first term yields slightly modified hypergeometric functions, which are well-behaved in the OPE limit $z \ra 0$. However, the second term arising from eq.~(\ref{eq:HyperIdentity}) will also have the prefactor $z^{2h_i - 2h_T}$, which is problematic. 

The scaling dimensions $h_i,h_T$ are independent and arbitrary. Even when they are integer spaced, the hypergeometric relation degenerates and one obtains a $\log z$ factor.  For $\ell < 2h_i - 2h_T$ this will not be a problem, but the action of a large number of derivatives will eventually yield terms which are singular in the OPE limit. Since we are specifically interested in the limit $\ell \ra \infty$, we will always encounter such divergent terms.

We therefore find that this alternate ``direct'' approach for extracting the $[TT]_{0,\ell}$ OPE coefficients cannot be applied to the $\Ocal_i$ conformal block alone, as its cross-channel OPE limit is not well-defined \cite{Petkou:1994ad, Hoffmann:2000tr, Hoffmann:2000mx}.
For this correlation function to be well-behaved in the OPE limit, there \emph{must} be additional operators which eliminate the $z\ra0$ branch cut in this single conformal block.  This is one reason that the bootstrap equation provides such a non-trivial constraint on CFTs.  More generally, in the OPE limit $T(z, \bar z) T(0)$ we obtain an expasion in $\frac{1}{z \bar z}$ with arbitrary power-law singularities (not necessarily integers) multiplied by an analytic function of $z, \bar z$, whose powers depend on the spin of operators in the OPE.  Individual conformal blocks do not have such an expansion in their cross-channels.\footnote{This behavior can easily be seen to generalize to conformal blocks in higher dimensions. As shown in appendix~\ref{sec:GlobalBlocks}, in $d$ dimensions the conformal block for a scalar primary $\Ocal_i$ has the OPE limit
\be
g_{\Ocal_i}(v,u) \approx \fr{\G(\De_i) \G(\De_i-\De_T)}{\G^2(\De_i-\fr{\De_T}{2})} \, (1-z)^{\fr{\De_i}{2}} \, \phantom{}_2 F_1 \left( \fr{\De_T}{2}, \fr{\De_T}{2}; \De_i-\fr{d}{2}+1; 1-z \right),
\ee
where we have specifically taken the limit $\bar{z} \ra 0$ with fixed $z$. We can then use eq.~(\ref{eq:HyperIdentity}) to rewrite the hypergeometric function, obtaining a term proportional to $z^{\De_i-\De_T-\fr{d}{2}+1}$. Individual conformal blocks in arbitrary dimensions therefore possess a $z \ra 0$ branch cut which renders the cross-channel OPE limit ill-defined.}

However, we know from the previous sections that the $\CO_i$ conformal block does contain enough information to determine the large $\ell$ behavior of the $[TT]_{0, \ell}$ OPE coefficients.  This suggests that one should be able to see that after  ``adding in'' the necessary conformal blocks required to make the cross-channel well-defined, the direct approach is insensitive at leading order in large $\ell$ to the details of how the extra conformal blocks were added.     We will see that Mellin space is well-suited for clarifying these issues.  


\subsection{Existence of the OPE and Boundedness of Mellin Amplitudes}
\label{sec:OPEandMellin}

We can obtain a new perspective on the cross-channel OPE limit by rewriting the correlation function in Mellin space \cite{Mack, JoaoMellin, NaturalLanguage, Paulos:2011ie}. The Mellin amplitude makes all OPE limits manifest, enabling a unified treatment of different channels.

The Mellin amplitude $\Mcal(\de_{ij})$ associated with a particular correlation function is defined by the relation
\be
\< \Ocal_1(x_1) \cdots \Ocal_n(x_n) \> = \int_{-i\infty}^{i \infty} [d\de_{ij}] \prod_{i < j} \left( \G(\de_{ij}) x_{ij}^{-2\de_{ij}} \right) \Mcal(\de_{ij}).
\ee
The Mellin variables $\de_{ij}$ are constrained to  the phase space 
\bq
\de_{ij} = \de_{ji}, \ \ \
\de_{ii} = 0, \ \ \
\sum_{j \neq i} \de_{ij} = \De_i,
\eq
which can be viewed as analogous to momentum conservation constraints on Mandelstam variables in scattering amplitudes.
For our four-point correlator, there are only two independent Mellin variables, which we can choose to be $s = \de_{14}$, $t = \de_{12}$. We then obtain the simplified expression 
\be
\< \Ocal_i(\infty) \Ocal_i(1) T(z,\bar{z}) T(0) \> = (z \bar z)^{\Delta_i - \Delta_T}  \int_{-i \infty}^{i \infty} \fr{ds \, dt}{(2 \pi i)^2} \frac{\Mcal(s,t) \prod_{i < j} \G(\de_{ij})   }{  (z \bar z)^{t} ((1-z)(1-\bar z))^{s}}.
\label{eq:MellinTransform}
\ee
An advantage of Mellin amplitudes is that their poles correspond to the scaling dimensions of exchanged operators. More specifically, the residue of a simple pole is the OPE coefficient of the associated operator, while double poles give perturbative anomalous dimensions.  We can use the tools of complex analysis to study the contributions to the correlation function.

We see immediately that in the limit from section \ref{sec:TTfromLightcone} where $z \to 1, \bar z \to 0$, the leading $z, \bar z$ dependence is
\be
(1-z)^{-s} \, {\bar z}^{\Delta_i-\Delta _T-t} .
\ee
In this limit, the correlator will be dominated by the poles of the Mellin integrand with the largest values of $s$ and $t$.  From the analysis of section \ref{sec:LargeSpinOPEfromBootstrap} we expect that the leading poles in $s,t$ will be responsible for the behavior of the $[TT]_{n,\ell}$ OPE coefficients.

A primary focus will be the Mellin amplitude for the $\CO_i T \to \Ocal_i \to \CO_i T$ conformal block in general $d$
\be
B_{\Ocal_i}(s,t) = e^{i\pi(1-\De_i)} \left( e^{i\pi(2\De_i+\De_T-d-2s)}-1 \right) \fr{\G(s-\fr{\De_T}{2})\G(s-\De_i-\fr{\De_T}{2}+\fr{d}{2})}{\G^2(s)},
\label{eq:MellinBlock}
\ee
where we have assumed that $\Ocal_i$ is a scalar operator.  The first gamma function introduces an infinite set of poles at $s=\fr{\De_T}{2}-n$. 
Looking at eq.~(\ref{eq:MellinTransform}), we see that this leads to a series of terms proportional to $v^{\half(\De_i + 2n)}$, which corresponds to the exchange of $\Ocal_i$ and its descendants. The second gamma function also contains a set of unphysical poles at $s=\De_i + \fr{\De_T}{2} + \fr{d}{2}-n$, which are eliminated by the zeroes of the oscillatory prefactor.

So in the $z \to 1, \bar z \to 0$ limit from section \ref{sec:TTfromLightcone}, the correlator is dominated by the poles of the Mellin amplitude at\footnote{We are assuming that $\Delta_i - \Delta_T > 0 $, interpreting $\CO_i$ as `heavy' and $T$ as `light'.}
\be
s = \frac{\Delta_T}{2},  \ \ \ t = \Delta_i - \Delta_T,
\ee
where the latter comes from the $\Gamma(\delta_{34}) = \Gamma(t - \Delta_i + \Delta_T)$ in the definition of the Mellin space integrand.  This pole determines the limiting behavior in equation (\ref{eq:LightConeforTT}).  Much of the utility of the Mellin formalism comes from our ability to connect the dominant contribution to the correlator with a single simple pole.  Other contributions to the correlator that do not affect the residue of this pole, and that do not supercede its importance, will not affect the $[TT]_{n,\ell}$ OPE coefficients.

We are interested in studying the $\< \CO_i(\infty) \CO_i(1) T(z) T(0) \>$ correlation function directly in the OPE limit $z, \bar z \to 0$, in order to extract OPE coefficients for individual $[TT]_{n, \ell}$ operators.  
As discussed in the previous section, the $\CO_i T \to \CO_i \to \CO_i T$ conformal block has a branch cut emanating from $z = 0$, which obstructs the OPE limit.  We would like to understand how this manifests in Mellin space.  If we study the residue of a pole in the Mellin integrand at $s=s_*$, we can expand in the OPE limit to find
\be
(1-z)^{-s_*} \approx 1 + (s_*) z + \frac{s_*(s_* + 1)}{2} z^2 + \cdots,
\ee
a result which is analytic near $z=0$. Looking at the integrand of eq.~(\ref{eq:MellinTransform}), we see that the $(1-z)$-dependence takes the form $(1-z)^{-s}$, such that any non-analyticity about $z=0$ must come from the behavior of the Mellin amplitude as $|s| \ra \infty$. The analytic structure of correlation functions near $z=0$ is determined by the poles of the Mellin integrand in $t$ and the asymptotic behavior of the Mellin amplitude as a function of $s$.

To see this more generally, note that the holomorphic dependence of the correlator is
\be
 z^{\Delta_i - \Delta_T} \int \fr{ds \, dt}{(2 \pi i)^2} \tilde \Mcal(s,t)   z^{-t} (1-z)^{-s},
\ee
where we combined $\Mcal(s,t)$ with the gamma functions to write the integrand as $\tilde \Mcal(s,t)$.  Localizing the integral at a pole $t = t_*$, we find
\be
 z^{\Delta_i - \Delta_T - t_*} \int \fr{ds }{2 \pi i} \tilde \Mcal(s)   \sum_{k=0}^\infty (s)_k \frac{z^k}{k!} ,
\ee
where $(s)_k = s(s+1) \cdots (s+k-1)$ is the Pochhammer symbol, which produces a degree $k$ polynomial in $s$.  Thus to obtain an analytic function of $z$ near $z=0$, the integral of $\tilde \Mcal(s,t)$ times any polynomial in $s$ should be well-defined, which means that
\be
\label{eq:StatementofMellinBound}
\lim_{s \to \pm i \infty} \Mcal(s,t) \, e^{-2\pi|s|} |s|^k = 0 . 
\ee
We provide a more detailed argument in appendix \ref{app:DetailsMellinAsymptotics}.  One might wonder why we have focused on $s$; in fact via crossing symmetry we can exchange $s$ with $t$ or $t+s$, and so the Mellin amplitude should have equivalently bounded asymptotic behavior in these variables.  

Furthermore, in general we expect that when we expand the CFT correlator about $z=0$, the resulting series will have radius of convergence $1$.  This follows because the OPE of $\CO_i(z) \CO_i(0)$ converges until we reach the operator $\CO_i(1)$ in the 4-pt correlator.   Writing $1-z = r e^{i \theta}$,  we see that $|z| < 1$ implies that $\theta$ lies in the range $( -\frac{\pi}{2}, \frac{\pi}{2} )$.  So for the integrals defining the correlator in terms of the Mellin amplitude to be well-defined, we must have
\be
\label{eq:StrongerMellinBound}
\lim_{s \to \pm i \infty} \Mcal(s,t) \, e^{-\frac{3}{2} \pi |s|}  = 0 
\ee
to guarantee the convergence of the OPE when $|z| < 1$.   AdS field theories satisfy a much stronger polynomial bound, but it would be interesting to understand if this exponential bound has an interpretation in terms of string theory in AdS.

If the Mellin integrand $\tilde \Mcal(s,t)$ is well-behaved at infinity, then one might attempt to derive sum rules for the correlation function from contour integrals of $\tilde \Mcal(s,t)$.  For example, we can evaluate the integrals over $s$ and $t$ in equation (\ref{eq:MellinTransform}) by closing the contour of integration to the left or the right of the imaginary axis.  Because there is no contribution to the correlator as $s, t \to \pm i \infty$, these are equivalent.  So the sum rules simply express the fact that when the correlator is expanded in one channel, it will agree with its analytic continuation from any other channel.  But the  individual conformal blocks do not have this property.

\begin{figure}
\begin{center}
\includegraphics[width=\textwidth]{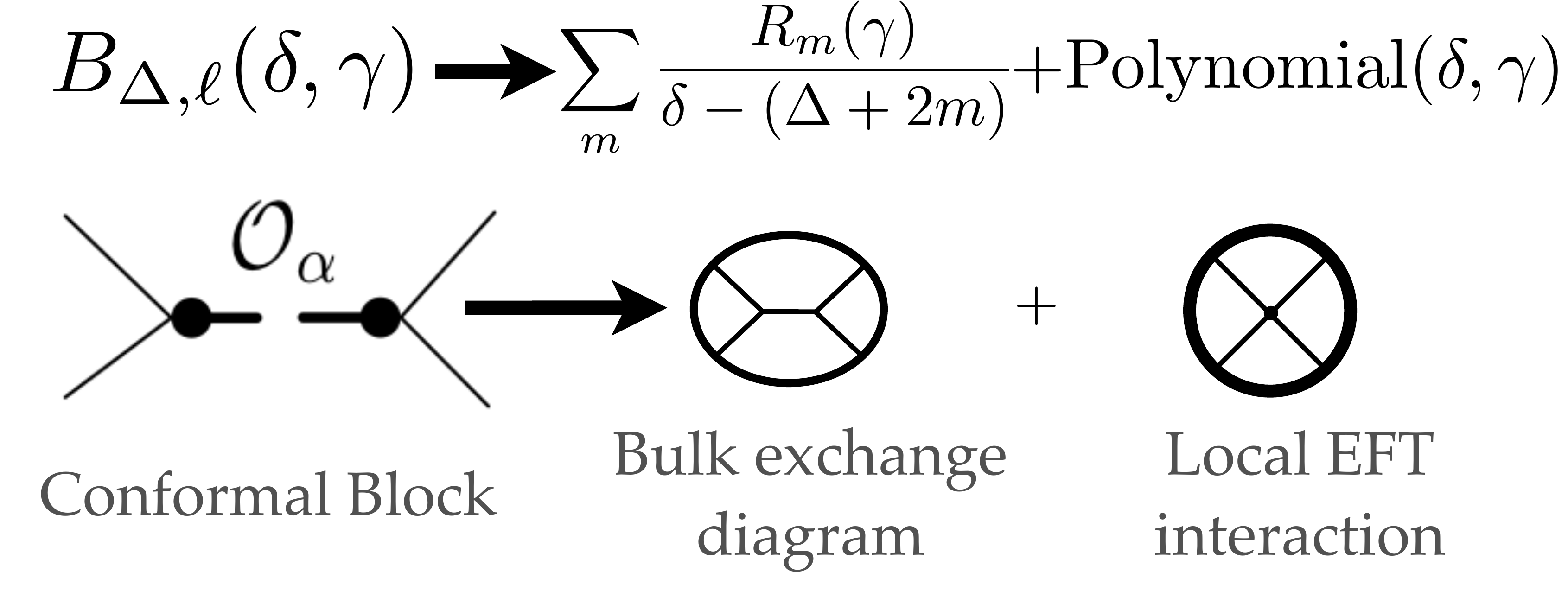}
\caption{This figure shows that neglecting the exponentially growing part of the Mellin amplitude for  conformal blocks with spin turns them into a Feynman diagram for an AdS exchange plus AdS contact interactions.  Demanding that all OPE limits are well-defined and expanding perturbatively in the $\<\CO_i \CO_i T\>$ OPE coefficients immediately suggests the existence of AdS Feynman diagrams  for the CFT correlators, as discussed in \cite{AdSfromCFT}.  }
\label{fig:BlocksToDiagrams}
\end{center}
\end{figure}

One can study the case where the OPE coefficient of $T$ in $\CO_i(x) \CO_i(0)$ is treated as a perturbative parameter, analogous to the `$1/N$' expansion, with other correlators taking the generalized free theory form (equivalent to a free field theory in AdS).  In that case, it can be shown that \cite{AdSfromCFT} the operators that cancel the bad asymptotic behavior of the Mellin amplitude must be $[\CO_i T]_{n, \ell}$ operators appearing in the $\CO_i(x) T(0)$ OPE.  The implied connection between conformal blocks and AdS Feynman diagrams is pictured in figure \ref{fig:BlocksToDiagrams}.


\subsection{Universality of Large Spin OPE Coefficients and a Theorem of Darboux}
\label{sec:UniversalityandDarboux}

We have seen that as long as the Mellin amplitude governing the $\< \CO_i(\infty) \CO_i(1) T(z, \bar z) T(0) \>$ correlator is bounded according to equation (\ref{eq:StatementofMellinBound}), all OPE limits will exist.  We can use this to show that the OPE coefficients of large spin  $[TT]_{n,\ell}$ with $\CO_i(x) \CO_i(0)$ are governed by operators in the $\CO_i(x) T(0)$ OPE with twist less than $\tau_i + \tau_T$, including $\CO_i$ itself.  

As discussed in section \ref{sec:OPElimit}, the OPE coefficients of $[TT]_{n, \ell}$ can be extracted by applying a differential operator $\Dcal_{n, \ell}$ at large $\ell$ and then taking the OPE limit $z \to 0$.  At large $\ell$ this requires the application of a large number of derivatives, and so the procedure is only sensitive to the large order expansion of the correlator in $z$.  A theorem of Darboux\footnote{See chapter VII of \cite{Dingle} for a discussion, \cite{Braaksma:1997:DTS:256442.256449} for rigorous statements and proofs, and \cite{Basar:2013eka} for an application to resurgence theory. } states that the dominant contribution to these large order terms is determined by singularities on the circle of convergence.  The OPE expansion converges for $|z| < 1$ and can only have singularities at $z = 1$, so as expected from  section \ref{sec:LargeSpinOPEfromBootstrap}, the large spin $[TT]_{n,\ell}$ OPE coefficients will be determined by the operators that produce singularities in the $\CO_i(1) T(z)$ OPE.  

In other words, a small subset of the Mellin amplitude poles determine the $[TT]_{n,\ell}$ OPE coefficients at large $\ell$.  These coefficients do not depend on the asymptotic behavior of the Mellin amplitude as long as it obeys the bound from equation (\ref{eq:StatementofMellinBound}).  By unitarity in the $\CO_i(x) T(0)$ OPE limit, these poles must have positive residue, and so they can never be eliminated by the inclusion of other operators.  The utility of the Mellin amplitude language is that it makes it easy to cleanly separate the poles, which determine the $[TT]_{n,\ell}$ OPE at large $\ell$, from the asymptotic behavior, which must satisfy a bound but is otherwise irrelevant for the large spin OPE coefficients we are studying.  But the Darboux theorem by itself justifies keeping only the singular terms from the $\CO_i(1) T(z)$ OPE when computing $[TT]_{n,\ell}$ OPE coefficients at large $\ell$.

We can now complete the analysis discussed in section \ref{sec:OPElimit} and use differential operators to derive the OPE coefficients of $[TT]_{n,\ell}$.  For simplicity we will take $n=0$, and study only the leading twist operators. The differential operator must be either entirely holomorphic and anti-holomorphic, such that we write $\Dcal_\ell = \Dcal_{\ell,z}$ or $\Dcal_{\ell,\bar{z}}$ when we are studying the minimal twist $[TT]_{0,\ell}$.  The differential operators take the form \cite{JoaoMellin, Unitarity} discussed in appendix \ref{sec:2dDecomposition}:
\be
\label{eq:DifferentialOperator}
\Dcal_{\ell,z} = \fr{1}{\Ncal_\ell} \sum_{k=0}^\ell \fr{(-1)^k}{k!(\ell-k)!\G(2 h_T+k) \G(2 h_T+\ell-k)} \partial_3^k \partial_4^{\ell-k},
\ee
with a similar expression for the antiholomorphic operator, and $\Ncal_\ell$ a normalization for the $[TT]_{0,\ell}$ operators. 
To determine the action of $\Dcal_\ell$, we need to evaluate expressions of the form 
\be
\partial_3^k \partial_4^{\ell-k} \left[ (1-z_3)^{-h_T}(1-z_4)^{-h_T} \, \phantom{}_2 F_1 \left( h_T, h_T; 2 h_i; 
\fr{1-z_3}{1-z_4} \right) \right].
\ee
We are interested in the large $\ell$ limit, which means taking $O(\ell)$ derivatives with respect to both $z_3$ and $z_4$, or equivalently, expanding this function in $z_3$ and $z_4$ to very high order.  By the Darboux theorem \cite{Dingle,Braaksma:1997:DTS:256442.256449}, the result will be governed by the singularity structure of the function on the circle bounding its radius of convergence.  The potential singularities are at $z_3, z_4 = 1$, and are dominated by the power-law prefactors.  If we send $z_4 \to 1$ with fixed $z_3$ we obtain a much weaker singularity, as can be seen from the hypergeometric identity
\be
\phantom{}_2F_1\left(h_T,h_T;2h_i;\fr{1-z_3}{1-z_4}\right) = \left( \fr{1-z_4}{z_3-z_4} \right)^{h_T} \, \phantom{}_2F_1\left(h_T,2h_i-h_T;2h_i;\fr{z_3-1}{z_3-z_4}\right),
\ee
and the fact that the last hypergeometric function behaves logarithmically near $1$.  This is a much weaker singularity than the simultaneous limit $z_3 = z_4 \to 1$.  Similarly, if we send $z_3 \to 1$ with fixed $z_4$ then we obtain a weakened singularity.  Thus the large order terms will be governed by
\bq
\begin{split}
\partial_3^k \partial_4^{\ell-k} &\left[ (1-z_3)^{-h_T}(1-z_4)^{-h_T} \, \phantom{}_2 F_1 \left( h_T, h_T; 2 h_i; 1 \right) \right] \\
&=(-1)^{\ell} \left( h_T \right)_k \left( h_T \right)_{\ell - k} 
\frac{\Gamma^2 \left( 2 h_i \right) \Gamma^2 \left( 2 h_i - 2 h_T \right)}{\Gamma^2 \left( 2 h_i - h_T \right)}.
\end{split}
\eq
Now we can sum over $k$ weighted by the coefficients in equation (\ref{eq:DifferentialOperator}) to obtain the $[TT]_{0,\ell}$ OPE coefficients 
\be
\frac{C^{(\Ocal_i \Ocal_i)}_{[TT]_{0,\ell}} }{C^{(TT)}_{[TT]_{0,\ell}}}
= \frac{ \Gamma^2 \left(2 h_T\right) \Gamma \left(\frac{\ell }{2}+h_T\right)  \Gamma \left(\frac{\ell +1}{2}\right) }{4^{ h_T} \Gamma^2 (h_T ) \Gamma \left(\frac{\ell }{2}+h_T+\frac{1}{2}\right) \Gamma \left(\frac{\ell }{2}+2 h_T\right)}
\times
\fr{ \G(2\bar{h}_i) \G(2\bar{h}_i-2\bar{h}_T)}{ \G^2(2\bar{h}_i-\bar{h}_T) },
\ee
where for simplicity we computed the ratio of the coefficient with the generalized free theory coefficients of $[TT]_{0,\ell}$ in the $T(x)T(0)$ OPE.  The result matches equation (\ref{eq:TTOPECoeff}) in the large $\ell$ limit, the only regime where either expression is valid. 

More generally, consider a pair of distinct operators $S$ and $T$, where both $S$ and $T$ appear in the OPE $\CO_i(x) \CO'(0)$ with $h_{\CO'} \leq h_i$.  Thus there exists a conformal block $\CO_i T \to \CO' \to \CO_i S$ contributing to the $\< \CO_i \CO_i T S \>$ correlator.  We can use the same methods to extract the large $\ell$ OPE coefficients of the operator $[ST]_{n,\ell}$ with $\CO_i(x) \CO_i(0)$.  If we restrict to the leading twist contributions, then we can allow all of the operators in this correlator to have general spin, expressed as a difference $h-\bar h$ between holomorphic and anti-holomorphic scaling dimension.   The treatment of general $S$ is useful because we can view $S$ itself as an operator like $[TT]_{n,\ell}$, enabling us to recursively determine general $[TT\cdots T]_\ell$ OPE coefficients at leading twist.  We find that 
\be
\label{eq:CTSGeneral}
 C_{[TS]_{0,\ell}}^{(\CO_i \CO_i)}  &=&  C^{(\CO_i T)}_{\CO'} C^{(\CO_i S)}_{\CO'}
\sqrt{ \frac{  (2h_T)_\ell   \left(h_i+h_S - h_{\CO'}  \right)_ \ell^2 }{\ell! (2 h_S)_\ell     \left( 2 h_S+2 h_T + \ell - 1 \right)_\ell   } }
\times \fr{ \G(2\bar{h}_{\CO'}) \G(2\bar{h}_i-\bar{h}_T - \bar{h}_S)}{ \G (\bar h_i + \bar h_{\CO'} - \bar h_T)  \G (\bar h_i + \bar h_{\CO'} - \bar h_S) }
\nn \\ &&
  \times {}_3F_2\left(-\ell ,1-\ell
   -2 h_S,h_{\CO'}-h_i+h_T;h_{\CO'}-\ell -h_i-h_S+1,2 h_T;1\right), 
\ee
where the result applies at very large $\ell$, and is due to the conformal block $\CO_i T \to \CO' \to \CO_i S$; the hypergeometric $_3F_2$ function simplifies when $h_T = h_S$.  

Using the more general differential operators or the conglomeration technique  of \cite{Unitarity}  one could compute the OPE coefficients with $n \neq 0$.  It would be especially interesting to study the combined contributions of an infinite tower of $\CO'$, as is encountered in 2d CFTs \cite{Fitzpatrick:2014vua, Fitzpatrick:2015zha}.  Operators such as $[ST]_{n, \ell}$ have computable anomalous dimensions that can also be expressed as a perturbation series in $1/\ell$ and $\log (\ell)$, so in future work, it will be important to understand how these affect the use of $\Dcal_\ell$ to compute OPE coefficients and the anomalous dimensions themselves.  We expect that there will be a `derivative relation' \cite{JP, Unitarity} for these large spin operators, so the perturbative OPE coefficients should be proportional to a certain normalized derivative acting on the anomalous dimensions.

We can use this result to provide an alternate derivation of the summed conformal block coefficient from $[T^m]_\ell$ exchange in equation (\ref{eq:FockSpaceBlockCoefficient}) for the case $m=3$.  We need to take $S = [TT]_{0,k}$ and combine it with $T$ to form a spin $\ell$ operator in all possible ways, so we have
\be
P_{\sum [TTT]_{0,\ell}}^{(11,22)} = \sum_{k=0}^\ell C_{[[TT]_k T]_{0,\ell-k}}^{(11)}  \times C_{[[TT]_k T]_{0,\ell-k}}^{(22)}   
\ee
for the minimal twist $[TTT]_\ell$ operators, where we take $\CO' = \CO_1$ and $\CO_2$ when we compute the first and second OPE coefficient, respectively.   We will not try to evaluate this expression in full generality, but instead we will treat the special case $h_T = 1$.  Then, displaying only the $\ell$ dependent factors in equation (\ref{eq:CTSGeneral}) for simplicity, the hypergeometric function simplifies and we obtain 
\be
P_{\sum [TTT]_\ell}^{(11,22)} &=& \sum_{k=0}^\ell \frac{\sqrt{\pi } (2 \ell -2 k + 3) \Gamma (\ell
   +3)}{2^{2\ell+3}(k+1) (\ell -k +1) (\ell -k +2) (2 \ell -k+4) \Gamma \left(\ell
   +\frac{5}{2}\right)}
\nn \\
& \approx & \frac{\sqrt{\pi} \log (\ell)}{2^{2\ell + 3} \ell^{\frac{3}{2}} }
\ee
at large $\ell$.  This matches the expectation from equation (\ref{eq:FockSpaceBlockCoefficient}), using equation (\ref{eq:DefinesGamma0}) for $\gamma_0$, once we take $h_T = 1$ and add the holomorphic and anti-holomorphic contributions.  Note that the $\log(\ell)$ dependence, which is crucial for eikonalization arises here because each Fock space primary $[[TT]_{k} T]_{\ell-k}$ has a conformal block coefficient proportional to $1/k$.

\section{Future Directions}

The primary motivations for this work were to study classical background fields in AdS purely from a CFT viewpoint, to extend and systematize the lightcone OPE limit and the corresponding $1/\ell$ perturbation theory, and to determine how much information about the CFT spectrum and OPE coefficients can be obtained from the existence of just a few operators and a single term in the OPE $\CO(x) \CO(0) \supset T$.   

The large spin expansion provides a concrete realization of an old hope for the bootstrap -- that OPE coefficients can be determined entirely from the CFT spectrum.  For example, if we know the dimensions of large spin operators of the form $[\CO \CO]_{n, \ell}$, then in principle we can immediately determine the OPE coefficients of $\CO(x) \CO(0)$ with all operators of twist less than $2 \tau_\CO$.  It is particularly interesting that as $\tau_\CO$ increases, we obtain more and more information about the OPE coefficients of $\CO$; perhaps in some situations one can make universal predictions about the properties of large dimension operators \cite{Jackson:2014nla}.  This also suggests that one might diagnose sub-AdS scale locality in the CFT by using heavy probe operators, without ever making an assumptions about a $1/N$ expansion.

 Now that we can compute the OPE coefficients of the Fock space operators $[TT\cdots T]_{\ell}$ from the OPE coefficients of their elementary consitituent, $T$, we can systematize the $1/\ell$ perturbation theory and attempt to analyze specific CFTs.  In future work it may be fruitful to use differential operators or integral conglomeration \cite{JoaoMellin, Unitarity} to directly extract OPE coefficients and anomalous dimensions from the singular parts of correlators.  

Recent work has shown that many universal features of quantum gravity in AdS$_3$, including the Hawking temperature of black holes, can be derived from the bootstrap \cite{Hartman:2014oaa, Fitzpatrick:2014vua} and Virasoro conformal blocks \cite{Fitzpatrick:2015zha} at large central charge, without other assumptions about the CFT data.  We would like to generalize these results as far as possible to CFTs in $d>2$ dimensions.  The universality of gravity suggests that these results should depend on only a few OPE coefficients and modest assumptions concerning the CFT spectrum.  For example, we would like to understand when high dimension operators behave as a thermal background for the correlators of light operators, and which CFT data determine the relationship between energy and temperature.  This will require incorporating the corrections to our results that produce non-linear classical fields in AdS, perhaps using an analysis that parallels \cite{Fitzpatrick:2015zha}.

Our results also support the idea that multiple weak interactions can build up to produce a large effect \cite{Camanho:2014apa}, without making direct reference to an AdS description, although a true eikonal limit in CFT \cite{Kaviraj:2015cxa} would make for a more decisive demonstration.  With the exchange of the large spin Fock space under control, it will be exciting to perform a study of the eikonal limit in general CFTs, although additional assumptions may be necessary.  Conversely, it may be possible to understand what `large' spin means, i.e.\ what values of $\ell$ are large?  For example, one might generalize the cross section bounds on scattering amplitudes to AdS/CFT, reformulating them as statements about the deviation of OPE coefficients from generalized free theory values.  Since CFTs in radial quantization are always gapped, a Froissart-type bound could exist for general theories, regulating the range and strength of AdS interactions.

We were able to bypass a technical obstruction in order to directly obtain OPE coefficients for large spin Fock space operators such as $[TT\cdots T]_\ell$, leading to the simpler computational methods of section \ref{sec:MellinBoundedness}.  Conceptually, our analysis was based on a Darboux-type \cite{Dingle, Braaksma:1997:DTS:256442.256449} argument that controls the large order behavior of the OPE.  Resurgence methods have been used to study the OPE in general QFTs \cite{Shifman:2014fra, Dunne:2015eoa}.  Our analysis suggests a controlled setting for studying the large order behavior of the OPE $\CO_L(x) \CO_L(1)$, namely in the background created by some heavy operators $\CO_H(\infty)\CO_H(0)$, which should appear thermal at large $\Delta_H$.  The large order behavior of the light operator OPE $\CO_L(x) \CO_L(1)$ should be governed by the singularity structure of the cross-channel $\CO_L(x) \CO_H(0)$ OPE.

\section*{Acknowledgements}

We would like to thank Rich Brower, Chris Brust, Aleksey Cherman, Liang Dai, Gerald Dunne, Ethan Dyer, Thomas Hartman, Shamit Kachru, Ami Katz, Zuhair Khandker,  Jo\~ao Penedones, Gustavo Marques Tavares, Leonardo Rastelli, David Simmons-Duffin, and Mithat Unsal  for discussions.  We also thank Jo\~ao Penedones for comments on the draft.  JK and JW are supported in part by NSF grant PHY-1316665 and by a Sloan Foundation fellowship.  ALF was partially supported by ERC grant BSMOXFORD no. 228169. MTW was supported by DOE grant DE-SC0010025.

\appendix

\section{Lightcone Formulation of Correlation Functions}
\label{sec:2dDecomposition}

In this appendix, we consider the decomposition of CFT$_d$ conformal blocks into representations of the collinear, or lightcone, subgroup, as used in \cite{KomargodskiZhiboedov,Braun:2003rp}. The lightcone OPE limit $z \ra 0$, $\bar{z} \ra 1$ then isolates those representations with lowest ``lightcone twist'', rewriting $d$-dimensional calculations in the simpler language of CFT$_2$.

\subsection{Lightcone Subgroup and Four-Point Functions}
In an arbitrary number of spacetime dimensions $d$, a combination of conformal transformations can reduce a 4-pt function to
\be
\< \Ocal_1(x_1) \Ocal_2(x_2) \Ocal_3(x_3) \Ocal_4(x_4) \> \ra \< \Ocal_1(\infty) \Ocal_2(1) \Ocal_3(x) \Ocal_4(0) \>.
\ee
The locations of three of the operators are therefore fixed, with only the two-dimensional location of the final operator (the points $0,1,x$ define a 2d plane) as the remaining degree of freedom.

Mirroring the setup in CFT$_2$, we can parametrize $x$ in terms of (anti)holomorphic coordinates $z,\bar{z}$. The conformal generators associated with these directions can then be combined into the suggestive form
\be
L_{-1} \equiv P_{z}, \quad L_{1} \equiv K_{z}, \quad L_0 \equiv \half (D+M_{z\bar{z}}),
\ee
with equivalent antiholomorphic generators. These generators form an $SL(2,\mathbb{C})/\mathbb{Z}_2$ ``lightcone'' subgroup of the full conformal group, with the commutation relations
\be
\comm{L_{\pm1}}{L_0} = \pm L_{\pm 1}, \quad \comm{L_1}{L_{-1}} = 2 L_0,
\ee
which matches the familiar 2d global conformal algebra.

The (anti)holomorphic generators $L_i,\bar{L}_j$ all commute with the conformal Casimir $C_d$. Irreducible representations of the full conformal group, associated with a primary operator with scaling dimension $\De$ and spin $\ell$, can therefore be broken up into irreps of the lightcone subgroup. These irreps are each characterized by their associated ``lightcone primary'', which is an eigenstate of $L_0,\bar{L}_0$ with eigenvalues $h,\bar{h}$.

We can start with the spin-$\ell$ primary state $|\Ocal_{\mu_1 \cdots \mu_\ell}\>$. This $SO(d)$ multiplet consists of various eigenstates of the angular momentum generator $M_{z\bar{z}} = L_0 - \bar{L}_0$, with corresponding eigenvalues
\be
m \equiv h - \bar{h} = \ell,\ell-1, \cdots,-\ell+1,-\ell.
\ee
All of these separate components are eigenstates of $D = L_0 + \bar{L}_0$ with the same scaling dimension $\De = h+\bar{h}$. The state $|\Ocal_{\mu_1\cdots\mu_\ell}\>$ is therefore divided into a set of distinct lightcone primary states $|\Ocal_{h,\bar{h}}\>$ with
\bq
\begin{matrix}
h & = & \fr{\tau}{2}+\ell, & \fr{\tau}{2}+\ell-1, & \cdots, & \fr{\tau}{2}+1, & \fr{\tau}{2}, \\
\bar{h} & = & \fr{\tau}{2}, & \fr{\tau}{2}+1, & \cdots, & \fr{\tau}{2}+\ell-1, & \fr{\tau}{2}+\ell,
\end{matrix}
\label{eq:SpinDecomposition}
\eq
where $\tau \equiv \De-\ell$ is the twist of the original primary operator.

Each of these lightcone primaries $\Ocal_{h,\bar{h}}$ defines an irrep of the lightcone subgroup, populated by the (anti)holomorphic descendants of the form $L_{-1}^m \bar{L}_{-1}^n |\Ocal_{h,\bar{h}}\>$. However, there are additional states created by acting with the transverse generators $P_\perp$. These states are also lightcone primaries,
\be
L_1 P_\perp |\Ocal_{h,\bar{h}}\> = \comm{L_1}{P_\perp}|\Ocal_{h,\bar{h}}\> = 0,
\ee
which have increased scaling dimension
\be
L_0 P_\perp |\Ocal_{h,\bar{h}}\> = \Big( P_\perp L_0 + \comm{L_0}{P_\perp} \Big)|\Ocal_{h,\bar{h}}\> = \left( h+\half \right) P_\perp|\Ocal_{h,\bar{h}}\>,
\ee
with matching antiholomorphic expressions. These states each define a new lightcone irrep, consisting of descendants of the form $L_{-1}^m \bar{L}_{-1}^n P_\perp^k |\Ocal_{h,\bar{h}}\>$.

A single representation of the full $d$-dimensional conformal group therefore generically decomposes into an infinite number of representations of the lightcone subgroup, with minimum ``twist'' $\tau$ and maximum ``spin'' $\ell$.

\subsection{Conformal Blocks and Lightcone OPE Limit}

Now that we understand the structure of the 2d lightcone subgroup, let's return to the original 4-pt correlation function. We'll first consider the case where all of the external operators are scalars, then later generalize to the case of arbitrary spin.

This correlation function can be expanded in terms of $d$-dimensional conformal blocks,
\be
\< \Ocal_1(\infty) \Ocal_2(1) \Ocal_3(z,\bar{z}) \Ocal_4(0) \> = (z\bar{z})^{-\half(\De_3+\De_4)} \sum_{\tau,\ell} P^{(12,34)}_{\tau,\ell} g^{(d)}_{\tau,\ell}(z,\bar{z}).
\ee
By taking the limit $z\ra0$, we can then isolate those conformal blocks with minimum twist $\tau_m$, which is of course bounded from below by unitarity.

Using the work of the previous section, we can then decompose each of these minimal twist conformal blocks in terms of an infinite number of lightcone irreps, parametrized by their associated (anti)holomorphic scaling dimensions,
\be
g^{(d)}_{\tau,\ell}(z,\bar{z}) = \sum_{h,\bar{h}} P_{h,\bar{h}} \, g^{(2)}_{h,\bar{h}}(z,\bar{z}),
\ee
where the coefficients $P_{h,\bar{h}}$ are completely fixed by conformal symmetry (and are therefore \emph{not} simply products of OPE coefficients). In the same small $z$ limit, this sum is dominated by those 2d conformal blocks with minimum holomorphic dimension $h$. In fact, there is only one such lightcone primary, with $h=\fr{\tau}{2}$ and $\bar{h} = \fr{\tau}{2}+\ell$.

In the lightcone OPE limit, we can therefore replace the minimal twist $d$-dimensional conformal blocks with the corresponding 2d blocks,
\be
g^{(d)}_{\tau,\ell}(z,\bar{z}) \ra g^{(2)}_{\tau,\ell}(z,\bar{z}) \qquad (z \ra 0).
\ee
We can then turn to the cross-channel expansion of our correlator,
\be
\< \Ocal_1(\infty) \Ocal_2(1) \Ocal_3(z,\bar{z}) \Ocal_4(0) \> = \big( (1-z)(1-\bar{z}) \big)^{-\half(\De_2+\De_3)} \sum_{\tau,\ell} P^{(14,23)}_{\tau,\ell} g^{(d)}_{\tau,\ell}(1-z,1-\bar{z}). \nn\\
\ee
Taking the limit $\bar{z} \ra 1$, we can again isolate the minimal twist conformal blocks. Each of these conformal blocks can then be decomposed into lightcone representations, with the dominant contribution coming from the one lightcone primary with $\bar{h}=\fr{\tau}{2}$ and $h = \fr{\tau}{2}+\ell$.

We therefore see that in the lightcone OPE limit $z\ra0,\bar{z}\ra1$ the bootstrap equation for $\<\Ocal_1\Ocal_2\Ocal_3\Ocal_4\>$ in arbitrary $d$ reduces to an effectively two-dimensional expression
\be
\sum_{\ell} P^{(12,34)}_{\tau_{m_1},\ell} g^{(2)}_{\tau_{m_1},\ell}(z,\bar{z}) \approx \fr{(z\bar{z})^{\half(\De_3+\De_4)}}{((1-z)(1-\bar{z}))^{\half(\De_2+\De_3)}} \sum_{\ell} P^{(14,23)}_{\tau_{m_2},\ell} g^{(2)}_{\tau_{m_2},\ell}(1-z,1-\bar{z}), \nn\\
\label{eq:2DLightconeBootstrap}
\ee
where $\tau_{m_1}, \tau_{m_2}$ are the minimal twists contributing to the $t$- and $s$-channels, respectively. This reduction to 2d conformal blocks makes manifest the $d$-independence of the asymptotic large $\ell$ results derived in \cite{Fitzpatrick:2012yx, KomargodskiZhiboedov}.

This decomposition can be easily generalized to correlation functions involving operators with nonzero spin. As shown in eq.~(\ref{eq:SpinDecomposition}), the components of each operator split into multiple lightcone primaries with $h,\bar{h}$ set by the twist $\tau$ and spin $\ell$ of the original operator. Correlation functions involving the operators $\Ocal_{\mu_1\cdots\mu_\ell}$ can then be written in terms of correlation functions built from the $\Ocal_{h,\bar{h}}$ operators, with the various correlators related by conformal transformations. These effectively 2d correlation function can then be studied using the conformal bootstrap, with the lightcone OPE limit taking a similar form to eq.~(\ref{eq:2DLightconeBootstrap}).

We therefore see that to determine the leading large $\ell$ behavior of minimal twist OPE coefficients and anomalous dimensions, which are studied in the lightcone OPE limit, it is sufficient to use two-dimensional conformal blocks, regardless of the number of spacetime dimensions.

\subsection{Differential Operators for OPE Limits in $d=2$}

Consider two operators $S_d$ and $T_d$ of general spin in $d$ dimensions.  If we pick a $2$-plane and decompose $S_d$ and $T_d$ according to the conformal group in this plane, for each of $S_d$ and $T_d$ we find operators $S, \bar S$ and $T, \bar T$ with minimal twist in the 2-plane.  For example, $S$ has $h_S = \frac{1}{2} \tau_{S_d} + \ell_{S_d}$ and $\bar h_S = \frac{1}{2} \tau_{S_d}$, whereas $\bar S$ has $h \leftrightarrow \bar h$.  Next we construct minimal twist large spin operators $[ST]_{\ell}(0)$ by acting with the holomorphic differential operator
\be
\Dcal_{\ell,z} = \fr{1}{\Ncal_\ell} \sum_{k=0}^\ell \fr{(-1)^k}{k!(\ell-k)!\G(2 h_T+k) \G(2 h_S+\ell-k)} \partial_{z_1}^k \partial_{z_2}^{\ell-k},
\ee
on $S(z_1) T(z_2)$, and then sending $z_1, z_2 \to 0$.  Similarly, starting with the mostly anti-holomorphic $\bar S$ and $\bar T$ and acting with an anti-holomorphic 
\be
\Dcal_{\ell, \bar z} = \fr{1}{\Ncal_\ell} \sum_{k=0}^\ell \fr{(-1)^k}{k!(\ell-k)!\G(2 \bar h_T+k) \G(2 \bar h_S+\ell-k)} \partial_{\bar z_1}^k \partial_{\bar z_2}^{\ell-k},
\ee
we can obtain $[\bar S \bar T]_{\ell}(0)$.  Note that in our notation, the spin of $[ST]_{\ell}$ is actually $\ell + \ell_{S_d} + \ell_{S_T}$.  Also note that $[ST]_{\ell}$ and $[\bar S \bar T]_{\ell}$ combine to form one even parity, $\tau = \tau_{S_d} + \tau_{T_d}$ operator.  Since the operators $[ST]_{\ell}$ have anomalous dimensions, when using these operators to extract OPE coefficients care must be taken to separate the two effects.  

The form of the differential operators can be easily justified by considering linear combinations of
\be
\sum_k a_k  \left( L_{-1}^k S \right) \left( L_{-1}^{\ell-k} T \right)
\ee
and demanding that this operator be primary, i.e.\ that it be annihilated by $L_1$.  This fixes the differential operators \cite{JoaoMellin, Unitarity} up to normalization.

The normalization factors $\Ncal_\ell$ can be obtained by acting with $\Dcal_{\ell, z}$ and $\Dcal_{\ell, \bar z}$ on the generalized free theory correlators 
\be
\<  T(z_1) T(z_2)  T(z_3) T(z_4)  \> = \frac{1}{z_{13}^{2h_T}z_{24}^{2h_T} \bar z_{13}^{2 \bar h_T} \bar z_{24}^{2 \bar h_T}} + \mathrm{permutations}
\ee
in order to obtain the 2-point function of $[ST]_{\ell}$ with itself.  Then 
\be
(\Ncal_\ell)^2 = \frac{4^\ell (2 h_S + 2 h_T +\ell - 1)_\ell}{\ell! \Gamma(2 h_S) \Gamma(2 h_T) \Gamma(2 h_S + \ell)  \Gamma(2 h_T + \ell)}
\ee
with our conventions, where we take $h_S = \frac{1}{2} \tau_{S_d} + \ell_{S_d}$ by definition in this formula.

\section{Scalar Conformal Blocks at Small $u$}
\label{sec:GlobalBlocks}

For any spacetime dimension $d$, the s-channel contribution to a generic correlation function $\<\Ocal_1 \Ocal_2 \Ocal_3 \Ocal_4\>$ from the global conformal block for a scalar operator $\Ocal$ with scaling dimension $\De$ can be written as the double sum 
\be
g_\Ocal(v,u) = \sum_{m,n \geq 0} \fr{(\fr{\De+\De_{12}}{2})_n (\fr{\De-\De_{34}}{2})_n}{n!(\De+1-\fr{d}{2})_n} \, \fr{(\fr{\De-\De_{12}}{2})_{n+m} (\fr{\De+\De_{34}}{2})_{n+m}}{m!(\De)_{2n+m}} \, v^{\fr{\De}{2}+n} (1-u)^m,
\ee
where $\De_{ij} \equiv \De_i - \De_j$. The sum over $m$ can be evaluated exactly, giving us the simpler expression
\bq
\begin{split}
g_\Ocal(v,u) = \sum_{n \geq 0} \Bigg\{ &\fr{(\fr{\De+\De_{12}}{2})_n (\fr{\De-\De_{12}}{2})_n (\fr{\De-\De_{34}}{2})_n (\fr{\De+\De_{34}}{2})_n}{n!(\De+1-\fr{d}{2})_n (\De)_{2n}} \\
& \otimes \, v^{\fr{\De}{2}+n} \phantom{}_2 F_1 \left( \fr{\De-\De_{12}}{2}+n, \fr{\De+\De_{34}}{2}+n; \De+2n; 1-u \right) \Bigg\}.
\end{split}
\eq
We are specifically interested in the behavior of this conformal block as $u \ra 0$. In this limit, we can use the identity
\be
\phantom{}_2 F_1 \left( \fr{\De-\De_{12}}{2}+n, \fr{\De+\De_{34}}{2}+n; \De+2n; 1 \right) = \fr{\G(\De+2n) \G(\fr{\De_{12}-\De_{34}}{2})}{\G(\fr{\De+\De_{12}}{2}+n) \G(\fr{\De-\De_{34}}{2}+n)},
\ee
to obtain
\be
g_\Ocal(v,0) = \fr{\G(\De) \G(\fr{\De_{12}-\De_{34}}{2})}{\G(\fr{\De+\De_{12}}{2}) \G(\fr{\De-\De_{34}}{2})} \sum_{n \geq 0} \fr{(\fr{\De-\De_{12}}{2})_n (\fr{\De+\De_{34}}{2})_n}{n!(\De+1-\fr{d}{2})_n} v^{\fr{\De}{2}+n}.
\ee
We can now evaluate the sum over $n$, with the resulting expression
\be
g_\Ocal(v,0) = \fr{\G(\De) \G(\fr{\De_{12}-\De_{34}}{2})}{\G(\fr{\De+\De_{12}}{2}) \G(\fr{\De-\De_{34}}{2})} \, v^{\fr{\De}{2}} \, \phantom{}_2 F_1 \left( \fr{\De-\De_{12}}{2}, \fr{\De+\De_{34}}{2}; \De+1-\fr{d}{2}; v \right).
\ee

As a simple check, let us compare this expression to some known results for even $d$. Starting with $d=2$, we have the general conformal block
\be
g^{(2)}_{\tau,\ell}(v,u) = k'_{\tau+2\ell}(1-z) k'_\tau(1-\bar{z}) + k'_\tau(1-z) k'_{\tau+2\ell}(1-\bar{z}),
\ee
where we have defined
\be
k'_{2\beta}(x) = x^{\beta} \, \phantom{}_2 F_1 \left(\beta-\half \De_{12}, \beta+\half \De_{34}; 2\beta; x\right).
\ee
For scalar operators this reduces to the simpler expression
\be
g^{(2)}_\Ocal(v,u) = k'_\De(1-z) k'_\De(1-\bar{z}).
\ee
The limit $u \ra 0$ at fixed $v$ is equivalent to taking $\bar{z} \ra 0$ with fixed $z$, which gives us the result
\be
g^{(2)}_\Ocal(v,0) = \fr{\G(\De) \G(\fr{\De_{12}-\De_{34}}{2})}{\G(\fr{\De+\De_{12}}{2}) \G(\fr{\De-\De_{34}}{2})} \, k'_\De(1-z).
\ee
Using the relation $z = 1-v$, we see that this precisely matches our general expression when $d=2$. Turning to $d=4$, we have the general s-channel conformal block
\be
g^{(4)}_{\tau,\ell}(v,u) = \fr{(1-z)(1-\bar{z})}{\bar{z}-z} \Big( k'_{\tau+2\ell} (1-z) k'_{\tau-2} (1-\bar{z}) - k'_{\tau-2} (1-z) k'_{\tau+2\ell} (1-\bar{z}) \Big), \,
\ee
which for a scalar operator reduces to
\be
g^{(4)}_\Ocal(v,u) = \fr{(1-z)(1-\bar{z})}{\bar{z}-z} \Big( k'_\De (1-z) k'_{\De-2} (1-\bar{z}) - k'_{\De-2} (1-z) k'_{\De} (1-\bar{z}) \Big).
\ee
If we again take the limit $\bar{z} \ra 0$ at fixed $z$, we then obtain
\be
g^{(4)}_\Ocal(v,0) = \fr{\G(\De) \G(\fr{\De_{12}-\De_{34}}{2})}{\G(\fr{\De+\De_{12}}{2}) \G(\fr{\De-\De_{34}}{2})} \fr{v}{1-v} \left( k'_{\De-2} (v) - \fr{(\fr{\De+\De_{12}}{2}-1) (\fr{\De-\De_{34}}{2}-1)}{(\De-1)(\De-2)} k'_\De (v) \right). \qquad
\ee
Using a combination of hypergeometric identities, we can rewrite this expression as
\be
g^{(4)}_\Ocal(v,0) = \fr{\G(\De) \G(\fr{\De_{12}-\De_{34}}{2})}{\G(\fr{\De+\De_{12}}{2}) \G(\fr{\De-\De_{34}}{2})} \, v^{\fr{\De}{2}} \, \phantom{}_2 F_1 \left( \fr{\De-\De_{12}}{2}, \fr{\De+\De_{34}}{2}; \De-1; v \right),
\ee
which again matches our general expression with $d=4$.

\section{Details of Mellin Asymptotics}
\label{app:DetailsMellinAsymptotics}

Let us study the connection between analyticity and Mellin amplitude asymptotics more carefully.   Consider the asymptotic behavior of the full integrand in eq.~(\ref{eq:MellinTransform}), specifically for the case of the $\Ocal_i$ conformal block. Looking at eq.~(\ref{eq:MellinBlock}), we see that this particular Mellin amplitude only depends on $s$, such that the integral over $t$ takes the simple form
\be
\int \fr{dt}{2\pi i} \, \G(t) \G(\De_T-\De_i+t) \G^2(\De_i - s - t) \, u^{-t}.
\ee
where $u = z \bar z$.
This integral can be evaluated by closing the contour of integration in the left half of the complex plane. The integrand has an infinite set of poles at both $t = -n$ and $t=\De_i-\De_T-n$, for all non-negative integers $n$, whose residues lead to increasing powers of $u$. However, we are specifically interested in the limit $u \ll 1$, such that we can focus on the lowest poles in both series. The resulting integral then takes the schematic form
\bq
\begin{split}
\int \fr{dt}{2\pi i} \, \G(t) &\G(\De_T-\De_i+t) \G^2(\De_i - s - t) \, u^{-t} \\
&\sim \Big( \G^2(\De_i-s) + \G^2(\De_T-s) u^{\De_T-\De_i} \Big) \Big( 1 + O(u) \Big),
\end{split}
\eq
where we have suppressed any constant coefficients to focus on the asymptotic scaling with respect to $s$ and $u$.

Now that we have an approximate form for the integration over $t$, we can then turn to the resulting $s$ integral,
\be
\int \fr{ds}{2\pi i} \, \G^2(s) \Big( \G^2(\De_i-s) + \G^2(\De_T-s) u^{\De_T-\De_i} \Big) v^{-s} \Mcal_{\Ocal_i}(s,t).
\ee
Using Sterling's approximation for gamma functions,
this Mellin amplitude then takes the asymptotic form
\be
\Mcal_{\Ocal_i}(s,t) \sim \left( e^{-2i\pi s}-1 \right) |s|^{\fr{d}{2}-\De_i-\De_T} \qquad (|s| \ra \infty),
\ee
where we have suppressed any constant coefficients to focus on the asymptotic scaling. This amplitude grows exponentially for $s \ra +i\infty$, but is power law suppressed for $s \ra -i\infty$.

To determine the asymptotic behavior of the full integrand, we also need to include the prefactor
\be
\G^2(s) \Big( \G^2(\De_i-s) + \G^2(\De_T-s) u^{\De_T-\De_i} \Big) \sim \fr{e^{-2\pi|s|}}{|s|^2} \Big( |s|^{2\De_i} + |s|^{2\De_T} u^{\De_T-\De_i} \Big),
\ee
where we have again suppressed constant coefficients. This product of gamma functions is therefore exponentially suppressed for large imaginary $s$, such that it perfectly cancels the exponential growth of the Mellin amplitude. Putting it all together, we obtain the asymptotic integrand
\bq
\begin{split}
\G^2(s) \Big( \G^2(\De_i-s) &+ \G^2(\De_T-s) u^{\De_T-\De_i} \Big) v^{-s} \Mcal_{\Ocal_i}(s,t) \\
& \sim |s|^{\fr{d}{2}-\De_i-\De_T-2} \Big( |s|^{2\De_i} + |s|^{2\De_T} u^{\De_T-\De_i} \Big) e^{-s \log v} ,
\end{split}
\eq
where we have taken the limit $s \ra +i\infty$ and assumed that $v$ is real and positive. We are specifically interested in the leading behavior of this conformal block in the limit $u \ra 0$ with $\De_i > \De_T$. In that case, this conformal block will be dominated by the second term, leading to the schematic integral
\be
u^{\De_T-\De_i} \int ds \, |s|^{\De_T-\De_i+\fr{d}{2}-2} e^{-s \log v} \sim u^{\De_T-\De_i} \Big( -\log v \Big)^{\De_i-\De_T-\fr{d}{2}+1}.
\ee
Finally, we need to include the overall prefactors of $u,v$ from eq.~(\ref{eq:MellinTransform}), obtaining
\be
g_{\Ocal_i}(v,u) \sim v^{\half(\De_i + \De_T)} \Big( -\log v \Big)^{\De_i-\De_T-\fr{d}{2}+1} \sim z^{\De_i-\De_T-\fr{d}{2}+1},
\label{eq:BlockNonAnalytic}
\ee
where we have taken the same limit as before, $\bar{z} \ll z \ll 1$. We therefore exactly reproduce the non-analytic term discovered in section~\ref{sec:OPElimit}.

The general lesson of this analysis is that \emph{non-analyticity} in position-space correlation functions arises from \emph{exponential growth} in the associated Mellin amplitudes. In this particular example, the Mellin integral for the $\Ocal_i$ conformal block only develops a branch cut at $z\ra0$ due to the term $e^{-2\pi is}$, which cancels the exponential suppression of the set of gamma functions. More generally, any Mellin amplitude which grows as
\be
\Mcal(\de_{ij}) \gtrsim e^{2\pi|\de_{ij}|} \qquad (|\de_{ij}| \ra \infty),
\ee
will lead to non-analyticities in the correlation function.

The demand that a correlation function possess a well-defined OPE limit then translates to a bound on the asymptotic behavior of the Mellin amplitude. In other words, analyticity near $z,\bar{z} = 0$ requires that Mellin amplitudes must be \emph{exponentially bounded} at large $\de_{ij}$.

We can make this bound more precise by considering a schematic Mellin amplitude with the asymptotic form
\be
\Mcal(s,t) \sim e^{2\pi|s|} |s|^{\fr{d}{2}-\De_i-\De_T-\alpha},
\ee
where $\alpha$ is an arbitrary positive number. Using this generic Mellin amplitude, we then obtain a contribution to the correlation function of the form
\be
\int ds \, \G^2(s) \G^2(\De_T-s) \, v^{-s} \Mcal(s,t) \sim z^{\De_i-\De_T-\fr{d}{2}+1+\alpha}.
\ee
We therefore see that for any value of $\alpha$, the correlator still possesses a branch cut at $z\ra0$.\footnote{For the case where the resulting exponent is a non-negative integer $n$, the correlation function still has a branch cut of the form $z^n \log z$.} The requirement of a well-defined OPE limit is then equivalent to the asymptotic bound
\be
\Mcal(s,t) \, e^{-2\pi|s|} |s|^n \ra 0 \qquad (|s| \ra \infty),
\ee
for \emph{any} integer $n$.

\section{Eikonalization with Interactions and Mixing in $d=2$}
\label{sec:2dResults}

In this appendix, we discuss two examples of the eikonalization of conformal blocks for 2d CFTs. We first consider the exchange of a conserved current $J$ in theories with a global $U(1)$ symmetry. The OPE coefficients of multi-trace operators built from $J$ are highly constrained, such that their conformal blocks automatically exponentiate in any kinematic limit. We then discuss the exchange of the stress-energy tensor $T$ in theories with large central charge $c$.  Generalizing the `direct method' results of \cite{Fitzpatrick:2014vua}, we demonstrate that corrections to the eikonal behavior in eq.~(\ref{eq:VirasoroExponentiate}) have a natural interpetation as the exponentiation of gravitational interactions in AdS$_3$.  We find some interesting cancellations in these calculations, where several complicated terms add up to something significantly simpler.  These cancellations also occur in more complicated examples.

\subsection{Exponentiation of Currents}

Consider a 2d CFT with an Abelian conserved current $J_\mu$. This current can be split into independent (anti)holomorphic components, $J(z)$ and $\bar{J}(\bar{z})$. Similar to the stress-energy tensor $T(z)$, the holomorphic current can then be expanded into modes,
\be
J(z) = \sum_m z^{m-1} J_{-m},
\ee
with a similar expansion for the antiholomorphic $\bar{J}(\bar{z})$. These modes obey the simple commutation relations
\be
\comm{J_m}{J_n} = k m \, \de_{m,-n},
\label{eq:CurrentAlgebra}
\ee
where $k$ is simply a normalization factor arising from the $J(z)$ two-point function. Similar to the stress-energy tensor, the vacuum is annihilated by the non-negative modes,
\be
J_m |0\> = 0 \quad (m \geq 0).
\ee

Based on the associated Ward identity, we can also derive the commutation relations with any charged primary operators $\Ocal_i$,
\be
\comm{J_m}{\Ocal_i(z)} = q_i \, z^m \Ocal_i(z),
\ee
where $q_i$ is the charge associated with $\Ocal_i$. Using these commutation relations, we can then determine the contribution of multi-$J$ exchange to the correlation function $\<\Ocal_1^\dagger\Ocal_1\Ocal_2\Ocal_2^\dagger\>$.

As a simple check of this approach, we'll first construct the global conformal block associated with the primary operator $J(z)$. Following the approach reviewed in \cite{Fitzpatrick:2014vua}, this conformal block can be constructed using a projection operator,
\be
g_J(z) = \fr{\< \Ocal_1^\dagger(\infty) \Ocal_1(1) [\Pcal_J] \Ocal_2(z) \Ocal_2^\dagger(0) \>}{\< \Ocal_1^\dagger(\infty) \Ocal_1(1) \>\< \Ocal_2(z) \Ocal_2^\dagger(0) \>},
\ee
where $\Pcal_J$ is formed from the set of ``one-photon'' states,
\be
\Pcal_J = \sum_{m=1}^\infty \fr{ J_{-m} |0\> \<0| J_m}{\<J_m J_{-m}\>}.
\ee

Based on the commutation relations above, we see that this basis is automatically orthogonal,
\be
\< J_m J_{-n} \> = km \de_{mn}.
\ee
We can then use the commutation relations above to derive the full expression for the $J(z)$ global block,\footnote{For notational simplicity, from now on we'll suppress the locations of the operators, though they will always correspond to the correlation function $\<\Ocal_1^\dagger(\infty)\Ocal_1(1)\Ocal_2(z)\Ocal_2^\dagger(0)\>$.}
\bq
\begin{split}
g_J(z) &= \sum_{m=1}^\infty \fr{\< \Ocal_1^\dagger \Ocal_1 J_{-m} \> \<J_m \Ocal_2 \Ocal_2^\dagger\>}{\<\Ocal_1^\dagger \Ocal_1\> \<J_m J_{-m}\> \<\Ocal_2 \Ocal_2^\dagger\>} = -\sum_{m=1}^\infty \fr{q_1 q_2}{km} z^m \\
&= \fr{q_1 q_2}{k} \log(1-z) = -\fr{q_1 q_2}{k} z \, \phantom{}_2 F_1(1,1;2;z).
\end{split}
\eq
We therefore obtain the correct form for the global block, with an overall conformal block coefficient of $P^{(11,22)}_J=-\fr{q_1 q_2}{k}$, indicating that this approach is correct.

We can then turn to the contribution of the double-trace operators $[JJ]_{n,\ell}$. Rather than break this contribution into individual conformal blocks, we'll consider the projection operator built out of all possible ``two-photon'' states,
\be
\Pcal_{JJ} = \sum_{m\geq n} \fr{J_{-m} J_{-n} |0\> \<0| J_n J_m}{\< J_n J_m J_{-m} J_{-n} \>}.
\ee
As we are considering an Abelian $U(1)$ global symmetry, this basis continues to be orthogonal, which we can confirm by computing the inner product
\be
\< J_m J_n J_{-p} J_{-q} \> = k^2mn (\de_{mp}\de_{nq} + \de_{mq}\de_{np}).
\ee
Using these results, we find that the full two-photon contribution takes the simple form
\be
\sum_{n,\ell} P^{(11,22)}_{[JJ]_{n,\ell}} g_{[JJ]_{n,\ell}}(z) = \fr{\< \Ocal_1^\dagger \Ocal_1 [\Pcal_{JJ}] \Ocal_2 \Ocal_2^\dagger \>}{\< \Ocal_1^\dagger \Ocal_1 \>\< \Ocal_2 \Ocal_2^\dagger \>} = \half \left( -\fr{q_1 q_2}{k} z \, \phantom{}_2 F_1(1,1;2;z) \right)^2.
\ee

We can then easily generalize this result to states with an arbitrary number of $J_m$ operators, obtaining the full contribution of ``photon exchange''
\be
\sum_{n} P^{(11,22)}_{[J^n]} g_{[J^n]}(z) &=& \sum_{\{m_i\}} \fr{\<\Ocal^\dagger_1 \Ocal_1 J_{-m_1} \cdots J_{-m_n} \> \< J_{m_n} \cdots J_{m_1} \Ocal_2 \Ocal_2^\dagger \>}{\< \Ocal^\dagger_1 \Ocal_1\> \< J_{m_n} \cdots J_{m_1} J_{-m_1} \cdots J_{-m_n} \> \< \Ocal_2 \Ocal_2^\dagger \>} \nn \\
&=& \sum_{n=0}^\infty \fr{1}{n!} \left( -\fr{q_1 q_2}{k} z \, \phantom{}_2 F_1(1,1;2;z) \right)^n = \exp \left[ -\fr{q_1 q_2}{k} z \, \phantom{}_2 F_1(1,1;2;z) \right] \nn \\
&=& (1-z)^{\fr{q_1 q_2}{k}}.
\ee
We therefore see that the Abelian nature of this global symmetry automatically leads to the exponentiation of multi-$J$ exchange. Note that this behavior is quite general, without any need to consider a particular kinematic limit or assume that the OPE coefficients are perturbatively small.

\subsection{Virasoro Blocks and Graviton Mixing}

As discussed in section~\ref{sec:VirasoroEikonalization}, in $d=2$ the contributions of all multi-trace operators built from the stress-energy tensor $T$ can be grouped together into a single function, called the identity Virasoro block. In \cite{Fitzpatrick:2014vua}, the identity Virasoro block for the 4-pt function $\<\Ocal_1 \Ocal_1 \Ocal_2 \Ocal_2\>$ was shown to have the approximate form
\be
\Vcal(z) \approx (1-z)^{h_2(\alpha-1)} \left( \fr{\alpha z}{1-(1-z)^{\alpha}} \right)^{2h_2},
\label{eq:VirasoroBlock}
\ee
where $\alpha = \sqrt{1-24\fr{h_1}{c}}$, and $h_1,h_2$ are the holomorphic scaling dimensions of $\Ocal_1,\Ocal_2$. This approximate form specifically holds in the semi-classical limit $c \ra \infty$ with fixed $\fr{h_1}{c}, h_2$, with a similar expression for the antiholomorphic block $\bar{\Vcal}(\bar{z})$. 

We can rewrite this Virasoro block in the suggestive form $\Vcal(z) = \exp[f(z)]$, with
\be
f(z) = h_2 \Big[ (\alpha-1) \log(1-z) + 2\log(\alpha z) - 2\log(1-(1-z)^\alpha) \Big].
\ee
While this form doesn't seem any more useful, we can gain more intuition by expanding it as a power series in $\fr{h_1}{c}$,
\be
f(z) \approx h_2 \left[ 2 \left(\fr{h_1}{c}\right) z^2 \phantom{}_2F_1(2,2;4;z) + O\left(\fr{h_1^2}{c^2} \right) \right].
\ee
To leading order, this exponent therefore matches the $T$ global conformal block! In other words, in the limit $c\ra\infty$ with fixed $\fr{h_1 h_2}{c}$, the identity Virasoro block is simply
\be
\Vcal(z) \approx \exp\Big[P_T \, g_T(z)\Big],
\ee
with $P_T = \fr{2h_1h_2}{c}$. We therefore see that to produce the leading contribution due to multi-$T$ exchange, one merely needs to calculate the contribution of ``one-graviton'' exchange,
\be
f(z) \approx \sum_{m=2}^\infty \fr{\<\Ocal_1 \Ocal_1 L_{-m}\> \<L_m \Ocal_2 \Ocal_2 \>}{\<\Ocal_1\Ocal_1\>\<L_m L_{-m}\>\<\Ocal_2\Ocal_2\>} = \fr{2h_1h_2}{c} z^2 \, \phantom{}_2 F_1 (2,2;4;z),
\ee
and exponentiate the result. Note that the modes $L_m$ are simply defined by the expansion
\be
T(z) = \sum_m z^{m-2} L_{-m}.
\ee

But what about the subleading corrections to the Virasoro block? Why does the full exchange of the stress-energy tensor not generically eikonalize like that of conserved currents? We can see the answer to this most clearly by considering the Virasoro commutation relations,
\be
\comm{L_m}{L_n} = (m-n) L_{m+n} + \fr{c}{12} m (m^2-1) \de_{m,-n}.
\label{eq:VirasoroAlgebra}
\ee
The second term in eq.~(\ref{eq:VirasoroAlgebra}) matches the structure of the $J$ commutation relations in eq.~(\ref{eq:CurrentAlgebra}). However, the first term corresponds to the fact that the OPE $T(x)T(0) \supset T$. This term then reflects the \emph{self-interactions} of gravitons in AdS$_3$.

In the limit $c\ra\infty$, the second term dominates, such that the algebra matches that of the Abelian current $J$. The calculation of multi-$T$ contributions then matches the work of the previous section, leading directly to eikonalization.

The corrections to this result at finite $c$ arise because this basis is \emph{not orthogonal}. Due to self-interactions, there is mixing between states with different numbers of ``gravitons''. The effects of this mixing are suppressed by factors of $1/c$, but must be included to obtain the full Virasoro block in eq.~(\ref{eq:VirasoroBlock}). We shall now argue that the subleading $O\left(\fr{h_1^k}{c^k}\right)$ corrections to $f(z)$ can be computed directly, with structure matching that of $k \ra 1$ graviton mixing.

We shall specifically consider the first correction to $f(z)$, which can be found by continuing to expand the expression to $O\left(\fr{h_1^2}{c^2}\right)$,
\be
\de f(z) \approx h_2 \left( \fr{h_1^2}{c^2} \right) \left( -\half \left( 2 z^2 \, \phantom{}_2 F_1 (2,2;4;z) \right)^2 - \fr{12}{5} z^3\log(1-z) \, \phantom{}_2 F_1 (3,3;6;z) \right).
\ee
There are only two possible contributions with the correct $h_1,h_2,c$-dependence to reproduce this expression. The first is the subleading corrections to ``two-graviton'' exchange, and the second is the contribution of ``two-to-one graviton mixing'', which arises due to the nonzero overlap
\be
\< L_{m+n} L_{-m} L_{-n} \> = \fr{c}{12} n(n^2-1) (2m+n).
\ee

Let's first start with the subleading $2\ra2$ corrections. Consider the contribution from some general two-graviton state,
\be
\fr{\< \Ocal_1 \Ocal_1 L_{-m} L_{-n}\>\<L_n L_m \Ocal_2 \Ocal_2\>}{\< \Ocal_1 \Ocal_1\> \<L_n L_m L_{-m} L_{-n}\> \<\Ocal_2 \Ocal_2\>} = \left( \fr{12 h_1}{c} \right)^2 \fr{(m-1)(n-1) h_2^2 + n(n-1) h_2}{(1+\de_{mn}) m (m+1) n(n+1)} z^{m+n}. \quad
\ee
The leading $h_2^2$ term simply corresponds to two copies of the one-graviton exchange, and is symmetric under $m \leftrightarrow n$. However, the second term has the correct $h_2$-dependence to potentially match $\de f(z)$, though it has a very asymmetric form,
\be
\de f^{(2\ra2)}_{m,n} = 144 h_2 \left( \fr{h_1^2}{c^2} \right) \fr{(n-1)z^{m+n}}{(1+\de_{mn}) m (m+1) (n+1)}.
\ee

However, the state $L_{-m} L_{-n}|0\>$ is not orthogonal to the one-graviton state, so the process of Gram-Schmidt forces us to also subtract the overlap with the one-particle state, leading to the $2\ra1$ mixing correction,
\bq
\begin{split}
\de f^{(2\ra1)}_{m,n} &= -\fr{\< \Ocal_1 \Ocal_1 L_{-m} L_{-n}\>\<L_n L_m L_{-m-n}\> \<L_{m+n} \Ocal_2 \Ocal_2\>}{\< \Ocal_1 \Ocal_1\> \<L_n L_m L_{-m} L_{-n}\> \<L_{m+n} L_{-m-n}\> \<\Ocal_2 \Ocal_2\>} \\
&= - \left( \fr{12 h_1}{c} \right)^2 \fr{(n-1) (2m+n) h_2 }{ (1+\de_{mn}) m(m+1) (m+n) (m+n+1)} z^{m+n}.
\end{split}
\eq
This expression has the same $h_2$-dependence as the subleading $2\ra2$ term, so we can combine them to obtain the much simpler expression
\be
\de f^{(2\ra2)}_{m,n} + \de f^{(2\ra1)}_{m,n} = 144 h_2 \left( \fr{h_1^2}{c^2} \right) \fr{(m-1) (n-1) z^{m+n}}{(1+\de_{mn}) (m+1)(n+1)(m+n)(m+n+1)}.
\ee
Note that, unlike the two individual pieces, this full expression is symmetric under $m \lra n$.

Let's now consider the full contribution from all two-graviton states. Including the corrections from both $2 \ra 2$ and $2 \ra 1$, we obtain
\bq
\begin{split}
\sum_{m \geq n} \Big( &\de f^{(2\ra2)}_{m,n} + \de f^{(2\ra1)}_{m,n} \Big) \\
&= h_2 \left( \fr{h_1^2}{c^2} \right) \left( -\half \left( 2 z^2 \, \phantom{}_2 F_1 (2,2;4;z) \right)^2 - \fr{12}{5} z^3\log(1-z) \, \phantom{}_2 F_1 (3,3;6;z) \right),
\end{split}
\eq
which \emph{precisely} matches the correction $\de f(z)$ obtained from the expansion of eq.~(\ref{eq:VirasoroBlock})!

It therefore appears that the semi-classical identity Virasoro block actually possesses a much simpler structure than would na\"{i}vely be expected. The combination of both $2 \ra 2$ and $2 \ra 1$ exchange into a highly simplified form suggests that there is more straightforward means of organizing this calculation \cite{Fitzpatrick:2015zha}, such that their combined contributions can be interpreted as simply the perturbative mixing between one- and two-graviton states. Similar simplifications appears at $O\left(\fr{h_1^3}{c^3}\right)$, suggesting the full semi-classical Virasoro block can be written as
\be
\Vcal(z) \approx \exp \Big[ (1 \ra 1) + (2 \ra 1) + (3 \ra 1) + \cdots \Big].
\ee
Understanding this ``generalized eikonalization'' structure more quantitatively would be an interesting direction for future research into the structure of both Virasoro blocks and gravitational interactions in general AdS$_{d+1}$.

\bibliographystyle{utphys}
\bibliography{EikonalBib}

\end{document}